\documentclass[pdflatex,sn-basic]{sn-jnl}

\usepackage{tikz}\usetikzlibrary{positioning, arrows.meta}
\usepackage{graphicx}
\usepackage{amsmath,amssymb,amsfonts}
\usepackage{xcolor}
\usepackage{color}
\usepackage{subcaption}
\usepackage{booktabs}
\usepackage{array}
\usepackage{orcidlink}
\definecolor{lightgray}{gray}{0.9}
\usepackage{tabularx}
\usepackage{xcolor}%
\usepackage{textcomp}%
\usepackage{manyfoot}%
\usepackage{booktabs}%
\usepackage{algorithm}%
\usepackage{algorithmicx}%
\usepackage{algpseudocode}%
\usepackage{listings}%
\usepackage{float}
\usepackage{balance}
\usepackage{enumitem}
\setlist{nosep}
\usepackage[normalem]{ulem}

\usepackage[title]{appendix}

\usepackage[T5]{fontenc}
\usepackage[utf8]{inputenc}

\DeclareTextSymbolDefault{\DH}{T1}

\usepackage{CJKutf8}
\newcommand{\cjkterm}[1]{\begin{CJK}{UTF8}{min}#1\end{CJK}}

\newcommand{\comment}[1]{\textcolor{black}{#1}}
\newcommand{\revise}[1]{\textcolor{black}{#1}}

\begin{document}


\title[Graphilosophy: Graph-Based Digital Humanities Computing with The Four Books]{Graphilosophy: Graph-Based Digital Humanities Computing with The Four Books}

\author[1,2]{\fnm{Minh-Thu} \sur{Do}\orcidlink{0009-0007-7129-6354}}\email{24C02018@student.hcmus.edu.vn}

\author[1,2]{\fnm{Quynh-Chau} \sur{Le-Tran}\orcidlink{0009-0008-8473-9713}}\email{24C02003@student.hcmus.edu.vn}

\author[1,2]{\fnm{Duc-Duy} \sur{Nguyen-Mai}\orcidlink{0009-0008-0783-5955}}\email{24C02006@student.hcmus.edu.vn}

\author[1,2]{\fnm{Thien-Trang} \sur{Nguyen}\orcidlink{0009-0002-3143-3725}}\email{24C02021@student.hcmus.edu.vn}

\author[1,2]{\fnm{Khanh-Duy} \sur{Le}\orcidlink{0000-0002-8297-5666}}\email{lkduy@fit.hcmus.edu.vn}

\author[1,2]{\fnm{Minh-Triet} \sur{Tran}\orcidlink{0000-0003-3046-3041}}\email{tmtriet@fit.hcmus.edu.vn}

\author[3]{\fnm{Tam V.} \sur{Nguyen}\orcidlink{0000-0003-0236-7992}}\email{tamnguyen@udayton.edu}

\author*[1,2]{\fnm{Trung-Nghia} \sur{Le}\orcidlink{0000-0002-7363-2610}}\email{ltnghia@fit.hcmus.edu.vn}

\affil[1]{\orgname{University of Science, VNU-HCM}, \orgaddress{\city{Ho Chi Minh City}, \country{Vietnam}}}

\affil[2]{\orgname{Vietnam National University - Ho Chi Minh}, \orgaddress{\city{Ho Chi Minh City}, \country{Vietnam}}}

\affil[3]{\orgname{University of Dayton}, \orgaddress{\city{Dayton}, \state{Ohio}, \country{United States}}}

\abstract{
\comment{The Four Books have shaped East Asian intellectual traditions, yet their multi-layered interpretive complexity limits their accessibility in the digital age. While traditional bilingual commentaries provide a vital pedagogical bridge, computational frameworks are needed to preserve and explore this wisdom.} This paper bridges AI and classical philosophy by introducing Graphilosophy, an ontology-guided, multi-layered knowledge graph framework for modeling and interpreting The Four Books. Integrating natural language processing, multilingual semantic embeddings, and humanistic analysis, the framework transforms a bilingual Chinese-Vietnamese corpus into an interpretively grounded resource. Graphilosophy encodes linguistic, conceptual, and interpretive relationships across interconnected layers, enabling cross-lingual retrieval and AI-assisted reasoning while explicitly preserving scholarly nuance and interpretive plurality. \comment{The system also enables non-expert users to trace the evolution of ethical concepts across borders and languages, ensuring that ancient wisdom remains a living resource for modern moral discourse rather than a static relic of the past.} Through an interactive interface, users can trace the evolution of ethical concepts across languages, ensuring ancient wisdom remains relevant for modern discourse. A preliminary user study suggests the system’s capacity to enhance conceptual understanding and cross-cultural learning. By linking algorithmic representation with ethical inquiry, this research exemplifies how AI can serve as a methodological bridge, accommodating the ambiguity of cultural heritage rather than reducing it to static data. \comment{The Source code and data are released at \url{https://github.com/ThuDoMinh1102/confucian-texts-knowledge-graph}.}
}


\maketitle

\section{Introduction}

The Four Books (\cjkterm{四書})\footnote{\url{https://en.wikipedia.org/wiki/Four_Books_and_Five_Classics}}, including The Great Learning (\cjkterm{大學}), The Doctrine of the Mean (\cjkterm{中庸}), The Analects of Confucius (\cjkterm{論語}), and The Works of Mencius (\cjkterm{孟子}), occupy a central place in East Asian intellectual and moral history. As the foundation of Confucian philosophy, these texts have shaped education, politics, and ethics across China, Vietnam, Korea, and Japan for over two millennia, while embodying enduring ideals of virtue and social harmony. Among many commentarial traditions surrounding The Four Books, Chinese-Vietnamese \texttt{\textit{Commentaries on The Four Books}} (\cjkterm{四書評解}) \citep{TuThuBinhGiai}, a widely recognized luminary in East Asian philosophy, is notable for its pedagogical clarity and bilingual structure. \comment{A scholar of Eastern philosophy, Ly Minh Tuan structured this work to integrate Classical Chinese text, transliteration, Vietnamese translation, and commentary, making the sages’ thought accessible to modern readers.} The work presents Classical Chinese texts with modern Vietnamese translations and notes, and its introduction underscores the continued relevance of Confucian virtues such as benevolence and altruism in modern life  \citep{TuThuBinhGiai}. This commentary thus helps bridge ancient Confucian ethics and contemporary moral concerns.

Despite their enduring influence, computational research on The Four Books and related commentaries remains scarce. Traditional digitization projects focus on text preservation and retrieval, rarely modeling the dynamic interpretive layers found in annotated works where commentary, translation, and source text interrelate. \comment{
These gaps are exacerbated by broader issues in digital humanities and cultural heritage preservation. AI-assisted translation introduces conceptual asymmetries; mapping terms like \textit{ren} (\cjkterm{仁}, benevolence) or \textit{li} (\cjkterm{禮}, ritual propriety) into modern languages risks diluting universal ethical ideals into culturally specific, localized interpretations. Privileging a single translation or commentary tradition in AI models can inadvertently amplify specific localized perspectives while marginalizing others, raising critical questions of interpretive authority and representational bias \citep{zhu2024}.}

\revise{Recent advances in AI-driven text analysis and knowledge graphs (KGs) have expanded how large cultural corpora can be organized and explored, supporting access and relational interpretation in digital humanities research \citep{de2009nlp, ferro2025novel, Haslhofer2018KGhum}. While general-purpose infrastructures offer broad coverage, domain-specific cultural heritage graphs better capture historical and conceptual complexity, yet scholarship emphasizes that such systems are sociotechnical constructs whose representational choices shape interpretive authority \citep{Suchanek2024YAGO, Vrandevcic2014Wikidata, Barzaghi2025CHADKG, Bai2023ChineseKG, drucker2020, klein2020, liu2012}. Recent work highlights pluralistic graph-based models as a response to these concerns, but Confucian classics such as The Four Books remain challenging due to linguistic concision, polysemy, and dense commentary traditions that resist stable or discrete computational representation \citep{yuan2025costume, foka2025}.}

\revise{To address this, we propose Graphilosophy, an ontology-guided, multi-layered KG framework for modeling The Four Books and their commentaries. Graphilosophy functions as an interpretive infrastructure that makes relationships among texts, translations, commentaries, speakers, and concepts explicit and navigable. Our system transforms \texttt{\textit{Commentaries on The Four Books}} \citep{TuThuBinhGiai}, which integrates the original Classical Chinese with a modern Vietnamese translation and pedagogically oriented commentar, into a structured, machine-readable dataset that supports semantic search, philosophical reasoning, and educational applications. We construct a multi-layered KG representation to model the intertextual relationships between doctrine and interpretation, an essential foundation for semantic understanding of Eastern philosophy. By externalizing interpretive structures instead of concealing them within opaque models, the system supports plural readings and reflexive engagement.}

\comment{Our system addresses representational bias through a scalable and explicitly pluralistic design that treats knowledge modeling as an evolving, interpretive process rather than a fixed technical structure.} \revise{Central to Graphilosophy is a modular KG that supports expansion across linguistic, interpretive, and philosophical dimensions, allowing multiple translations and expert commentaries to coexist and reducing linguistic bias and singular interpretive authority. Its layered and extensible ontology separates textual, linguistic, conceptual, and commentary dimensions so each can evolve independently while remaining connected, accommodating ambiguity and multiplicity as core features of Confucian philosophy. This design addresses concerns that reductive representational models flatten philosophical nuance and reproduce power imbalances \citep{drucker2017, klein2020, drucker2020}, aligning technical scalability with interpretive values central to digital humanities practice.}

\comment{This paper addresses two interrelated questions concerning the design and implications of AI-mediated knowledge representations for classical philosophical texts. First, how can an ontology-guided, multi-layered knowledge graph framework support the representation of the semantic, interpretive, and translational plurality inherent in The Four Books, while making visible the asymmetries of language, conceptual framing, and interpretive authority embedded in their transmission? Second, to what extent can such a system meaningfully support philosophical learning and cultural preservation without reducing the openness, ambiguity, and historical situatedness of the original texts, and what challenges arise when attempting to align scalability and bias mitigation with the ethical and interpretive demands of this domain?}

Our contributions are as follows:
\begin{itemize}
    \item \revise{We assemble a digitally annotated corpus of The Four Books that integrates Classical Chinese texts, Vietnamese translations, and pedagogical commentaries, making visible the layered and mediated nature of meaning across languages and interpretations.}
    
    \item \revise{We propose a domain-specific KG that models textual, conceptual, and commentary relationships as interconnected layers, supporting interpretive plurality rather than fixed semantic closure.}
    
    \item \revise{We develop exploratory and visual tools that enable navigation across these layers, supporting semantic exploration, teaching, and interpretive inquiry in digital humanities contexts.}
    
    \item \revise{Through experiments and a user study, we show how AI-based representations can assist learning and research while preserving the central role of human interpretation in engaging with Confucian philosophical texts.}

    \item \comment{The Source code and data are released at \url{https://github.com/ThuDoMinh1102/confucian-texts-knowledge-graph}.}
\end{itemize}

\section{Related Work}

Digital humanities initiatives increasingly rely on computational approaches to preserve cultural heritage, utilizing large-scale platforms (Europeana\footnote{\url{https://www.europeana.eu/}}, the Perseus Digital Library\footnote{\url{https://www.perseus.tufts.edu/hopper/}}, the Chinese Text Project\footnote{\url{https://ctext.org/}}) that provide structured, multilingual access to historical materials. While these efforts effectively treat humanities texts as data to broaden access, they primarily emphasize digitization, search, and metadata organization. Consequently, they offer limited means to engage with the profound semantic, philosophical, and linguistic complexity that shapes classical works and their traditions of interpretation.

Similarly, Natural Language Processing (NLP) has become essential for engaging with premodern corpora \citep{Haslhofer2018KGhum}, adapting techniques to handle the interpretive challenges of archaic texts \citep{johnson2021classical} and utilizing multilingual models to enable large-scale cross-lingual alignment \citep{zhang-etal-2023-miracl}. However, these computational developments remain heavily focused on surface-level linguistic processing and information retrieval. They provide limited support for modeling the rich intertextual and interpretive relationships through which historical and philosophical meaning is actually produced. 

To capture this relational meaning, Knowledge Graphs (KGs) are increasingly adopted to structure cultural knowledge \citep{Zheng2024KnowledgeMining, Cui2024Ontology}. Yet, general-purpose KGs often privilege bibliographic structure over interpretive depth \citep{kokash2024brill}. From a digital humanities and sociotechnical perspective, this emphasis risks treating cultural texts as stable, objective data points rather than dynamic sites of ongoing interpretation. For Confucian classics, linguistic concision and dense commentarial traditions expose fundamental tensions between current AI representations and the interpretive complexity of the domain.

In response, our study advances a domain-specific, ontology-guided KG that reflects the layered, dialogical nature of Confucian philosophy. By situating multilingual NLP within this framework, we treat language technologies not as neutral ends in themselves, but as mediating infrastructures. Utilizing recent advances in graph analysis and multilingual semantic representation, this approach foregrounds relational meaning, interpretive plurality, and the ethical responsibilities of AI-mediated knowledge representation.

\section{Proposed Dataset}

To enable computational analysis, we construct The Four Books as a Classical Chinese–Vietnamese resource organized at three levels: (1) tri-parallel alignment incorporating phonetic bridging between the two languages, (2) dictionary-level lexical mapping for precise semantic correspondence, and (3) chapter-based exegesis capturing interpretive commentary and contextual meaning. Together, these components form an extended corpus that encodes the semantic, phonetic, and interpretive dimensions of Classical Chinese and Vietnamese, reflecting the linguistic depth and contextual sophistication of each chapter.

\subsection{Dataset Construction}

\subsubsection{Data Source}

The dataset consists of authoritative digital editions of The Four Books, including the original Classical Chinese texts and standard commentaries. We use Commentaries on The Four Books (\cjkterm{四書評解}) \citep{TuThuBinhGiai} as the primary source for Chinese–Vietnamese translations and exegetical notes, chosen for its comprehensive coverage, reliable bilingual annotation, and pedagogical depth. All materials were drawn from open-access repositories and cross-checked against printed editions to ensure textual accuracy.

\revise{The dataset is structured around three interrelated components, Main Text, Dictionary, and Expert Analysis, which together reflect the layered and mediated nature of meaning in Confucian traditions. The \textit{\textbf{Main Text}} consists of 2,222 sentences from The Four Books, organized according to their original textual hierarchy and presented in Classical Chinese alongside Sino-Vietnamese readings and modern Vietnamese translations, thereby preserving both linguistic structure and translational mediation. This component serves as the interpretive foundation of the dataset, enabling close engagement with source passages while situating them within broader pedagogical and semantic contexts. The \textit{\textbf{Dictionary}} includes 5,344 entries of Classical Chinese characters that were consolidated into 2,788 unique entries to account for polysemy and contextual variation, treating lexical ambiguity as an interpretive condition rather than a defect to be eliminated. The \textit{\textbf{Expert Analysis}} comprises 80 commentary entries authored by Ly Minh Tuan, providing pedagogical and interpretive perspectives that situate the texts within established traditions of explanation and learning. The separation of these components is an intentional design choice that supports transparency and interpretive plurality, allowing textual content, lexical interpretation, and commentary to evolve independently while remaining connected to the source texts, and framing computational structure as a mediating practice through which cultural knowledge is represented and transmitted.}

\subsubsection{Construction Pipeline}

\revise{The dataset construction followed a staged workflow that treats corpus preparation as an interpretive practice. It involved Preprocessing, Alignment, Refinement, and Structuring, with attention to preserving textual integrity and contextual meaning. During \textit{\textbf{Preprocessing}}, scanned sources were transformed into structured digital text, corrected, and normalized to remove layout artifacts using \comment{PaddleOCR\footnote{\url{https://www.paddleocr.ai/}}}, segmented according to the original textual logic, and enriched with contextual information such as provenance and authorship. Lexical materials were extracted and contextualized using metadata (book and chapter) to reflect the role of vocabulary and polysemy in shaping meaning.}

\revise{The \textit{\textbf{Strategy Alignment}} explicitly differentiated among three corpus types: the Tri-Parallel Corpus, which aligns Classical Chinese passages with Sino-Vietnamese readings and modern Vietnamese translations; the Lexical Corpus, which connects dictionary entries to their textual contexts to address polysemy and usage; and the Exegesis Corpus, which links expert commentaries to corresponding canonical passages. Alignment combined automated procedures with expert review, underscoring that relational mapping across languages and interpretations is an inherently interpretive process.}

\comment{Following automated processing, the \textit{\textbf{Refinement}} stage treats correction and verification as acts of interpretive responsibility rather than purely technical cleanup. First, heuristic recovery was used to identify and correct digitization artifacts that could distort meaning or disrupt textual continuity. This was followed by a close manual audit carried out by domain experts, who reviewed all aligned textual units to ensure coherence and fidelity across the Classical Chinese, Sino-Vietnamese, and modern Vietnamese layers. Rather than prioritizing mechanical alignment alone, this process foregrounds accountability, transparency, and human judgment in the mediation of cultural texts, establishing a reliable and ethically grounded foundation for the subsequent multi-layered knowledge representation.}

\revise{In the \textit{\textbf{Structuring}} stage, all materials were organized into interoperable formats with consistent identifiers, enabling the corpus to function as an integrated whole. This structure supports analysis, teaching, and exploration while maintaining transparency in how texts, translations, and interpretations are connected.}

\subsection{Dataset Description}

The finalized dataset comprises 2,222 segmented sentences, 80 scholarly commentary 
annotations, and a refined lexical dictionary of 2,788 entries, including 2,562 unique Classical Chinese characters and 23 core Confucian concepts
. These materials are instantiated in an ontology-driven KG containing 16,468 nodes and 71,249 edges spanning textual, linguistic, and interpretive layers
.

Rather than serving as a purely technical resource, this structure is designed to externalize how meaning is mediated through translation, commentary, and conceptual 
categorization in classical philosophy. The dataset is organized into three complementary components, including the \textit{Tri-Parallel Corpus}, the \textit{Lexical 
Dictionary}, and the \textit{Exegesis Corpus}, each preserving a distinct dimension of interpretive plurality: translational mediation, lexical polysemy, and scholarly 
commentary, respectively. Together, they form an interoperable cultural dataset that links text, translation, lexicon, and expert interpretation through a unified ontology, enabling AI systems to \textit{mediate}, rather than resolve, the complexity of classical philosophical knowledge.

\section{Methodology}

\subsection{Overview}

Our proposed Graphilosophy is a comprehensive, multi-stage framework that integrates advanced text processing, layered KG construction, and an interactive web interface powered by Google Gemini \citep{Google2023Gemini} for natural language querying. Our KG was built using the NetworkX library \citep{Hagberg2008NetworkX} in a modular manner. Rather than generating a monolithic graph, our system constructs six distinct yet interlinked layers, such as Textual, Linguistic, Conceptual, Commentary, Speaker, and Semantic (Figure~\ref{fig:ontology_layers}). Each layer is processed independently to preserve its unique analytical function and data integrity before being interconnected through ontology-guided relationships. This modular design maintains interpretive clarity within each dimension while enabling a unified, multi-perspective understanding of The Four Books and their accompanying commentaries.

\subsection{Ontology Design and Construction}

The Graphilosophy framework employs a custom multi-layered ontology specifically designed to model the bilingual Classical Chinese--Vietnamese corpus of \emph{The Four Books}. The ontology systematically structures textual, linguistic, conceptual, and interpretive dimensions across six interconnected layers (Meta, Textual, Linguistic, Conceptual, Commentary \& Speaker, and Semantic), comprising 20 entity classes and 18 directed relationship types. This design supports multi-hop reasoning, cross-lingual retrieval, and interpretive plurality while maintaining modularity and scalability.

\begin{figure}[!t]
\centering 
    \includegraphics[trim={0 0 0 3.3cm},clip, width=0.9\linewidth]{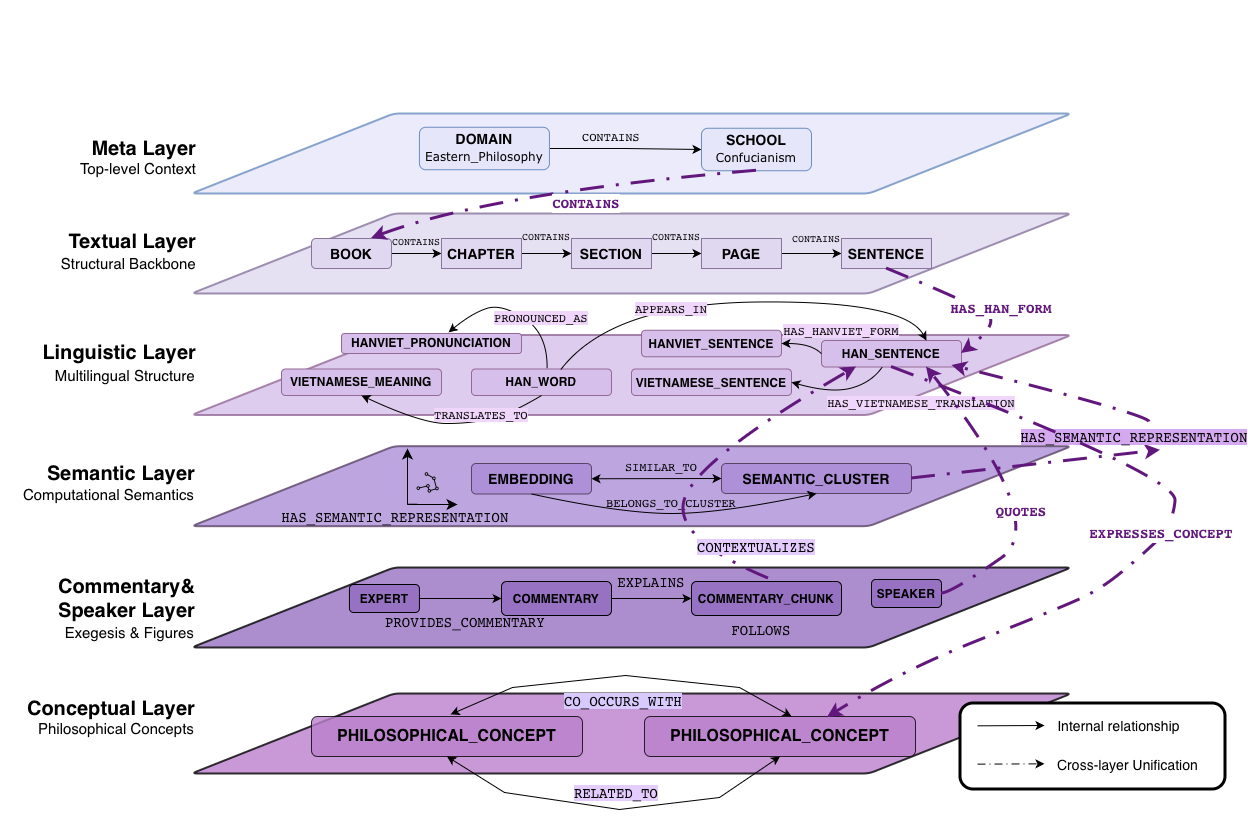}
\caption{The multi-layered ontology architecture of the Graphilosophy knowledge graph. The schema models the corpus across six distinct but interconnected layers to preserve structural, linguistic, and interpretive dimensions. Solid lines indicate intra-layer relationships, while dashed lines represent cross-layer unifications that enable complex, multi-hop reasoning.}
\label{fig:ontology_layers}
\end{figure}

Figure~\ref{fig:ontology_layers} illustrates the ontology architecture, depicting the sequential flow from the Meta Layer through the Textual, Linguistic, Conceptual, and Commentary \& Speaker Layers to the Semantic Layer, together with explicit cross-layer connections that unify the entire knowledge graph.

Table~\ref{tab:ontology} provides a comprehensive summary of all entity classes and relationship types, grouped by layer.

\begin{table}[!t]
\centering
\renewcommand{\arraystretch}{1.55} 
\caption{Entity classes and relationship types in the Graphilosophy KG ontology}
\label{tab:ontology}
\begin{tabularx}{\textwidth}{@{} 
    l 
    l 
    >{\raggedright\arraybackslash\hsize=1.3\hsize}X 
    >{\raggedright\arraybackslash\hsize=0.7\hsize}X 
@{}}
\toprule
\textbf{Layer} & \textbf{Type} & \textbf{Name} & \textbf{Method} \\
\midrule
\textbf{Meta}                     & Entity   & DOMAIN, SCHOOL & Predefined \\
\textbf{Textual}                  & Entity   & BOOK, CHAPTER, SECTION, PAGE, SENTENCE & Auto \\
\textbf{Textual}                  & Relation & CONTAINS, FOLLOWS, APPEARS\_IN & Auto \\
\textbf{Linguistic}               & Entity   & HAN\_SENTENCE, HANVIET\_SENTENCE, VIETNAMESE\_SENTENCE & Auto \\
\textbf{Linguistic}               & Entity   & HAN\_WORD, HANVIET\_PRONUNCIATION, VIETNAMESE\_MEANING & Auto \\
\textbf{Linguistic}               & Relation & HAS\_HAN\_FORM, HAS\_HANVIET\_FORM, HAS\_VIETNAMESE\_\allowbreak TRANSLATION, TRANSLATES\_TO, PRONOUNCED\_AS & Auto \\
\textbf{Conceptual}               & Entity   & PHILOSOPHICAL\_\allowbreak CONCEPT (Pattern matching) & Auto \\
\textbf{Conceptual}               & Relation & EXPRESSES\_CONCEPT, RELATED\_TO, CO\_OCCURS\_WITH & Auto + semi-manual \\
\textbf{Commentary \& Speaker}    & Entity   & EXPERT, COMMENTARY, COMMENTARY\_\allowbreak CHUNK, SPEAKER & Manual / Pattern detection \\
\textbf{Commentary \& Speaker}    & Relation & PROVIDES\_\allowbreak COMMENTARY, EXPLAINS, CONTEXTUALIZES, QUOTES & Manual + similarity $>$ 0.75 \\
\textbf{Semantic}                 & Entity   & EMBEDDING, SEMANTIC\_CLUSTER & Auto \\
\textbf{Semantic}                 & Relation & SIMILAR\_TO, BELONGS\_TO\_CLUSTER, HAS\_SEMANTIC\_REP & Auto \\
\bottomrule
\end{tabularx}
\vspace{0.7em}
\begin{small}
\textbf{Note}: Total of 20 entity classes and 18 relationship types.  \\
\textbf{Method}: Auto = fully automatic (rule-based); Semi = algorithm + human verification; Manual = expert-defined.
\end{small}
\end{table}

Relations are generated through three distinct methods (fully automatic rule-based, semi-automatic with human validation, and fully manual expert-defined) and unified into a single directed graph via explicit cross-layer links (e.g., \texttt{HAS\_HAN\_FORM}, \texttt{CONTEXTUALIZES}, \texttt{EXPRESSES\_CONCEPT}). This modular, ontology-guided structure externalizes interpretive processes, preserves semantic ambiguity inherent in Confucian philosophy, and provides a robust foundation for semantic search, philosophical reasoning, and educational exploration.

\revise{\subsection{Linguistic Processing}
\label{sec:linguistic_processing}
Classical Chinese presents unique computational challenges due to its brevity, 
polysemy, and context-dependence.
\paragraph{Polysemy Resolution via Contextual Embedding}
When a Classical Chinese character has multiple possible meanings, the system adopts a context-sensitive matching approach, using Multilingual-e5-large embeddings and cosine similarity, rather than enforcing a single dictionary definition. Meanings are evaluated in relation to the surrounding passage to support interpretive coherence, without treating the result as definitive. This design reflects digital humanities commitments to preserving semantic ambiguity and aligns with concerns about limiting algorithmic authority in the interpretation of philosophical texts. Example (\cjkterm{道} (\textit{Đạo})): This character has three primary meanings in our dictionary: path/road, to speak/say, and doctrine/The Way. In the passage “\cjkterm{子曰}:\cjkterm{參乎}! \cjkterm{吾道} \cjkterm{道一以貫之}” (Analects 4.15), the system interprets \cjkterm{道} in relation to the surrounding philosophical discourse, foregrounding its doctrinal sense based on the philosophical context of Confucius addressing his disciple about his unifying principle.
\paragraph{Segmentation in Context}
While Classical Chinese lacks punctuation, the system uses Ly Minh Tuan’s scholarly edition \citep{TuThuBinhGiai} as an interpretive reference rather than a neutral ground truth, and validates segment boundaries through cross-layer consistency with Sino-Vietnamese and modern Vietnamese translations. Ambiguous cases are resolved by favoring readings supported by both translation alignment and established commentary, while explicitly acknowledging the role of scholarly judgment. This design aligns with digital humanities principles of interpretive transparency and with concerns about making human assumptions visible in computational text processing.
\paragraph{Handling Untranslatable Concepts}
For philosophical concepts that cannot be fully captured in modern Vietnamese, the system preserves the original Classical Chinese term, links it to a broader conceptual category, and supplements it with expert commentary. For example, \cjkterm{仁} (\textit{Nhân/Ren}) is represented through multiple Vietnamese approximations and contextualized as a core virtue through scholarly explanation, rather than reduced to a single translation. This layered representation treats untranslatability as an interpretive feature, aligning with digital humanities emphases on semantic plurality and concerns about avoiding reductive representations of ethical concepts.
}

\paragraph{Success and Failure Cases}
\textbf{Success Cases:}
\begin{itemize}
    \item \textit{\cjkterm{人} vs.\ \cjkterm{仁} Disambiguation}: Despite identical Sino-Vietnamese 
    pronunciation ``\textit{Nhân}'', the system correctly distinguished \cjkterm{人} (human/person) 
    from \cjkterm{仁} (benevolence/virtue) in 98\% of cases (2,178/2,222 sentences). The 
    E5-Large embeddings position these characters in distinct semantic regions:  \cjkterm{人} clusters with pronouns and social roles, while \cjkterm{仁} clusters with virtue terms.  
    \item \textit{Cross-lingual Concept Retrieval}: A Vietnamese query for 
    ``\textit{đạo hiếu}'' (filial piety) successfully retrieved 47 Classical Chinese 
    sentences containing \cjkterm{孝} (\textit{hiếu}), even when the word ``\textit{đạo}'' was absent from 
    the original text, demonstrating effective semantic bridging.
    \item \textit{Speaker Attribution}: The pattern-based detection correctly 
    attributed 89\% of quotations to speakers (Confucius: ``\cjkterm{子曰}''; Mencius: 
    ``\cjkterm{孟子曰}''; disciples: ``\cjkterm{曾子曰},'' ``\cjkterm{子貢曰}'').
\end{itemize}
\textbf{Failure Cases:}
\begin{itemize}
    \item \textit{\cjkterm{樂} Phonetic Ambiguity}: The character \cjkterm{樂} can be read as 
    \textit{Lạc} (joy/pleasure) or \textit{Nhạc} (music). In sentences like ``\cjkterm{知之者},\cjkterm{不如好之者};\cjkterm{好之者},\cjkterm{不如樂之者}'' (Analects 6.20), the system occasionally confuses these readings when sentence structure is sparse. This phonetic ambiguity remains a challenging case where embedding-based semantic disambiguation alone may be insufficient without additional phonetic or syntactic cues.
    \item \textit{Implicit Subject Recovery}: Classical Chinese frequently omits subjects. In ``\cjkterm{學而時習之},\cjkterm{不亦說乎}'' (Analects 1.1), the implicit subject ``one who learns'' is not explicitly represented, limiting certain speaker-attribution queries.
\end{itemize}


These failure cases are not isolated edge cases but structurally predictable outcomes of embedding-based disambiguation applied to a language where phonetic 
and semantic information are not always recoverable from context alone; their systematic treatment is discussed in Section~\ref{sec:error_analysis}.

\revise{
\subsection{Semantic Chunking}
This workflow segments classical texts and commentaries based on philosophical continuity rather than fixed boundaries, treating segmentation as an interpretive intervention to preserve contextual integrity and navigate Classical Chinese polysemy. Linguistically, it integrates Classical Chinese, Sino-Vietnamese, and modern Vietnamese via a consolidated lexical resource that accommodates character-level ambiguity. By identifying core Confucian concepts through established taxonomies and contextual cues, the system explicitly supports multiple interpretations. Prioritizing this interpretive plurality over strict technical optimization limits algorithmic authority, ensuring responsible computational mediation of historical texts.
}

\begin{figure}[!t]
    \centering
    \includegraphics[trim={0 0 0 13cm},clip, width=\linewidth]{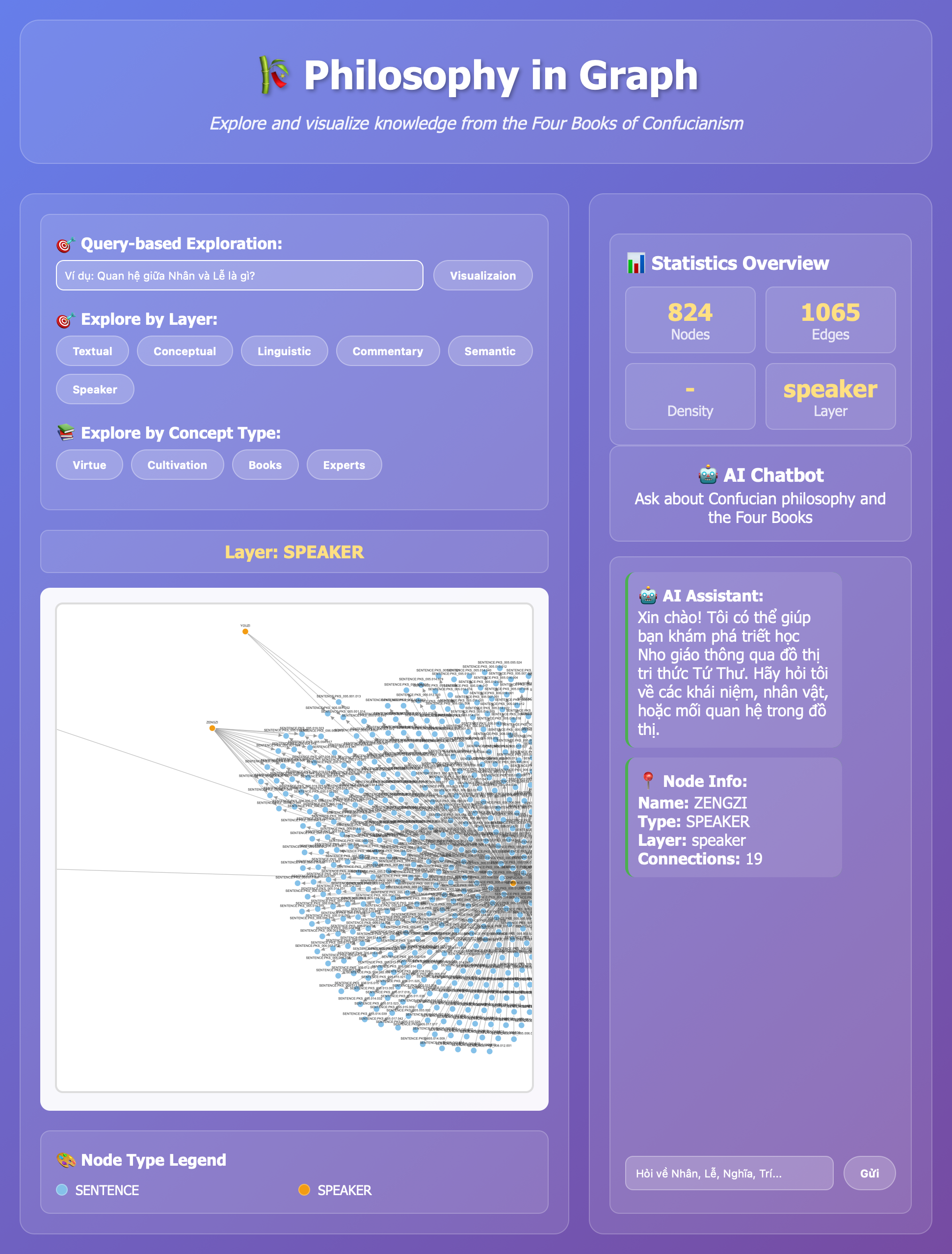}
    \caption{User interface of the proposed system integrating layered KG visualization, semantic search, and commentary exploration. The interface enables cross-lingual retrieval and interpretive analysis of The Four Books through Gemini-powered natural language querying. Some instructions in the interface are displayed in Vietnamese to enhance usability for local users.}
    \label{fig:interface}
\end{figure}

\subsection{Interactive System Interface}

The system provides an interactive platform for exploring, visualizing, and querying the KG via Google Gemini (Figure~\ref{fig:interface}), which interprets natural-language queries and orchestrates a hybrid search mechanism combining structured graph traversal with exact textual matching. Rather than generating responses in isolation, Gemini operates over explicitly defined KG context: natural-language queries are resolved into localized graph neighborhoods to maintain interpretive clarity, while verbatim inputs trigger direct text matching to ensure philological accuracy. This design promotes transparent and verifiable interaction with classical texts, supporting close reading, comparative analysis, and pedagogical use.

\section{Structural Evaluation}

\subsection{Core Graph Metrics and Sparsity}
\label{sec:graph_analysis}

This unified KG is a highly complex and structured data system, comprising 16,468 nodes, which are interconnected by 71,249 edges representing meaningful relationships between them. Despite the large number of nodes and edges, the network exhibits an extremely low density of 0.000263, classifying it as a sparse network. This sparsity demonstrates that connections are selectively generated to represent meaningful, non-trivial semantic relationships, making the graph efficient for targeted queries.

\comment{
\subsection{Knowledge Graph Structural Validation}
The internal consistency of the KG serves as a secondary validation of the segmentation quality. The graph currently hosts 16,468 nodes and 71,249 edges. A key metric is the \texttt{APPEARS\_IN} relationship, which constitutes 41.3\% of all edges ($29,417$ instances). This high density of connections between the 1,723 unique Classical Chinese words and the 2,222 segmented sentences confirms that the segmentation logic accurately mapped fine-grained linguistic units into their correct structural contexts.
}

\begin{figure}[t!]
    \centering
    \includegraphics[width=0.65\linewidth]{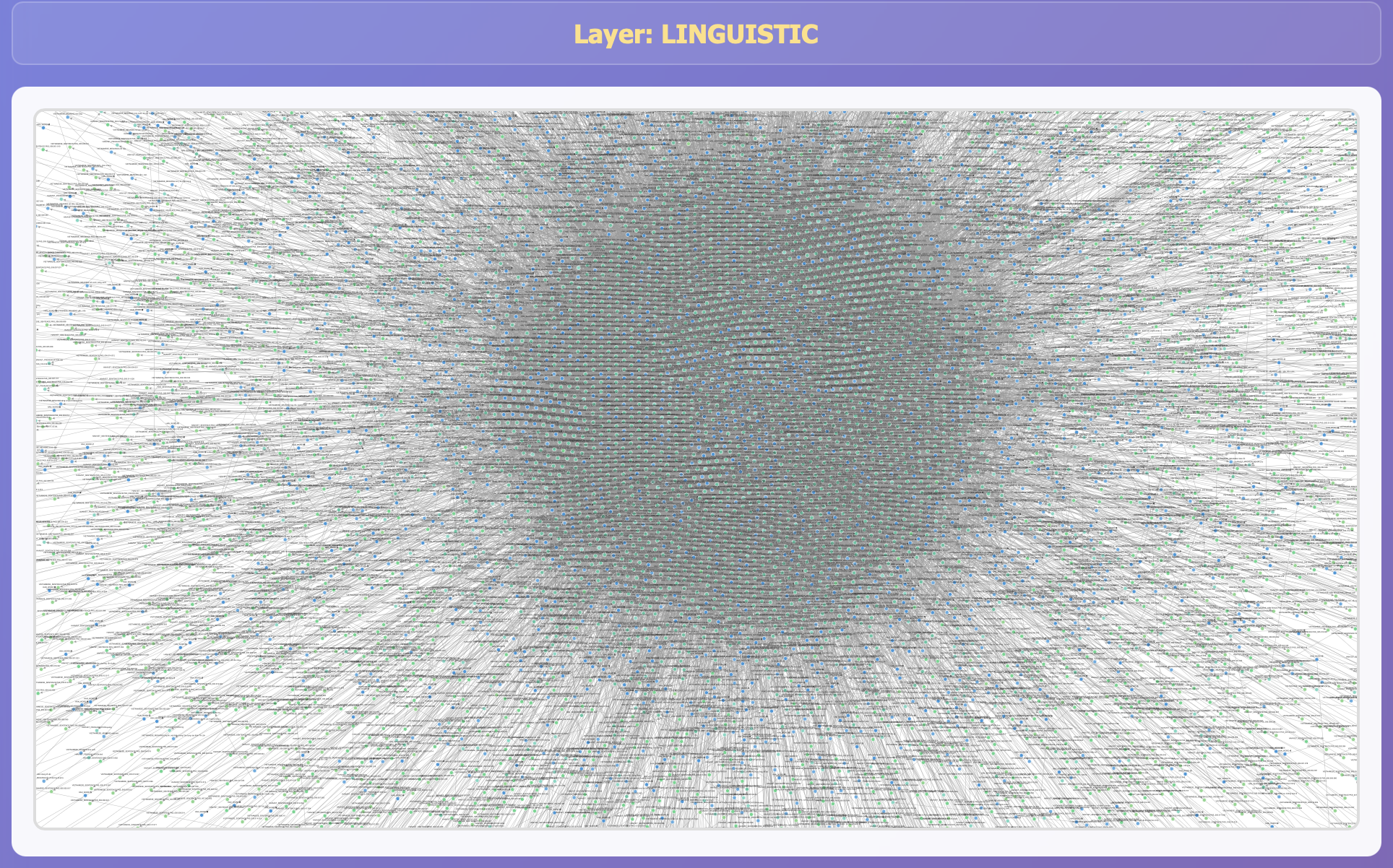}
    \caption{The high density of the Linguistic Layer reflects the substantial volume of nodes and edges within this layer, establishing a robust foundation for subsequent semantic analysis.}
    \label{fig:linguistic_density}
\end{figure}

\begin{figure}[t!]
    \centering
    \includegraphics[trim={0 4.5cm 0 0},clip, width=0.65\columnwidth]{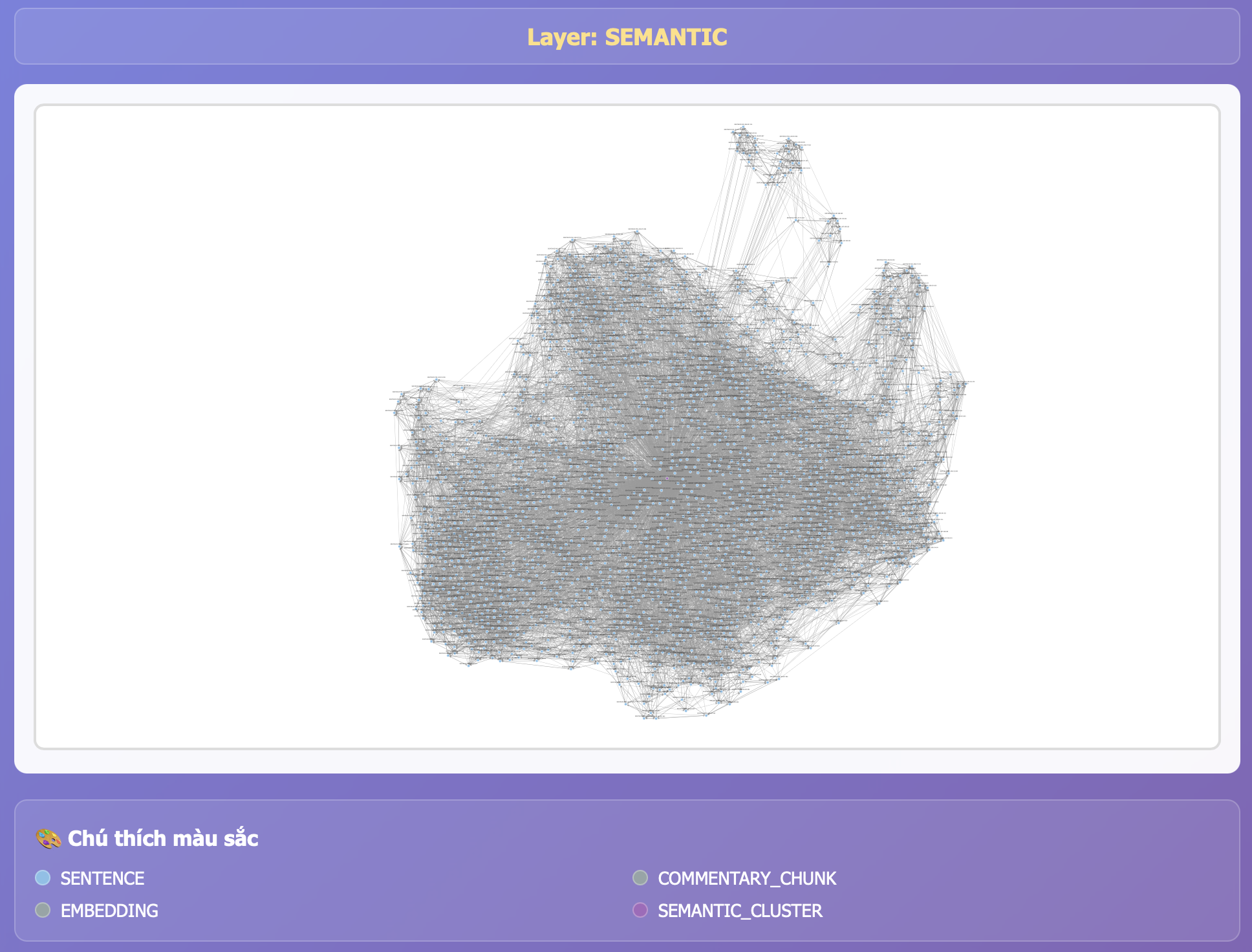}
        \caption{Semantic Layer structure. Visualization of the dense network formed by EMBEDDING nodes and SEMANTIC\_CLUSTERs, confirming the reliance on the semantic layer for similarity-based retrieval.}
        \label{fig:semantic_network}
        \vspace{-5mm}
\end{figure}

\begin{figure}[t!]
    \centering
    \begin{subfigure}[b]{0.48\textwidth}
        \centering
        \includegraphics[width=\textwidth]{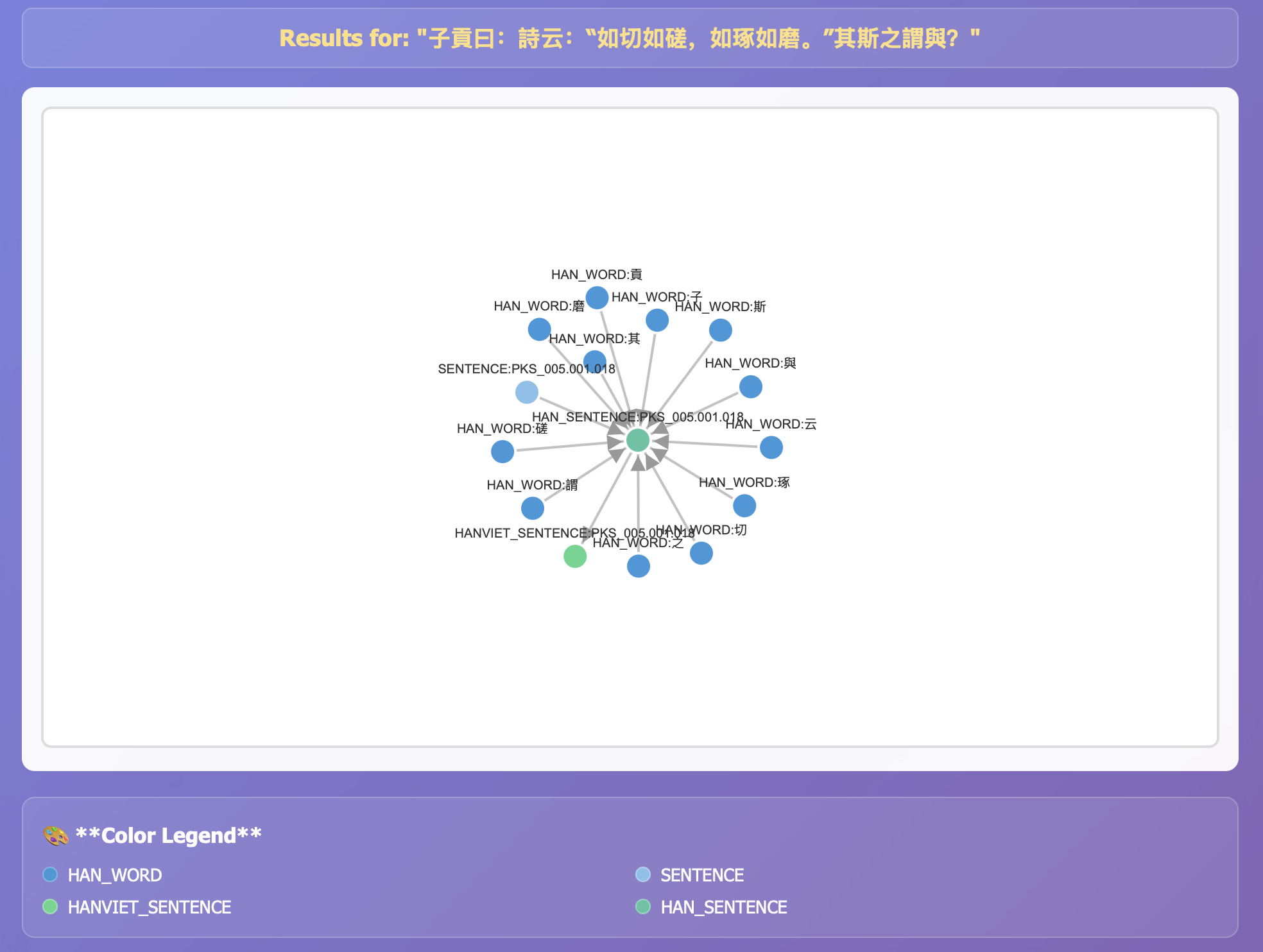}
        \caption{Focused subgraph for exact Classical Chinese query: ``\cjkterm{曾子曰}:\cjkterm{吾日三省吾身}...'' 
        (Analects 1.4). The central HAN\_SENTENCE node connects to constituent HAN\_WORD nodes, 
        with trilingual alignment to HANVIET\_SENTENCE and SENTENCE nodes.}
        \label{fig:search_focused}
    \end{subfigure}
    \hfill
    \begin{subfigure}[b]{0.48\textwidth}
        \centering
        \includegraphics[width=\textwidth]{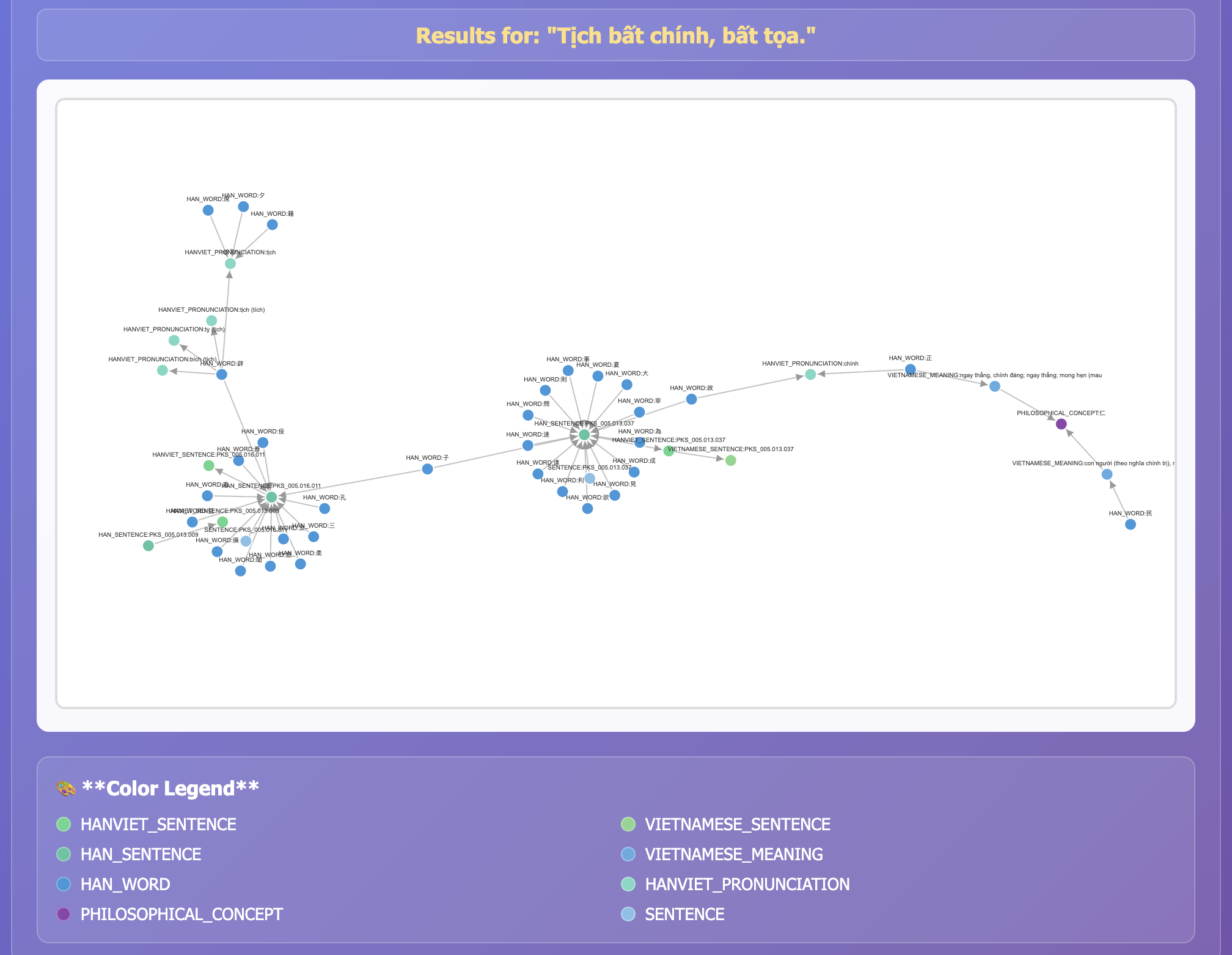}
        \caption{Multi-cluster subgraph for Vietnamese semantic query: ``Tịch bất chính, bất tọa.'' 
        Query retrieves multiple related passages, revealing cross-layer connections including 
        PHILOSOPHICAL\_CONCEPT nodes (magenta) and VIETNAMESE\_MEANING nodes (light blue).}
        \label{fig:search_multiclusters}
    \end{subfigure}
    \caption{\comment{Query-based focused visualization illustrating the BFS-based (depth = 1) search mechanism through Gemini model. 
    Unlike full-layer visualizations (Figures~\ref{fig:linguistic_density}, \ref{fig:semantic_network}) 
    which display the entire graph structure, these focused subgraphs present only the immediate neighborhood of query-matched nodes, reducing visual complexity while preserving interpretive context. (a) An exact single-sentence query produces a star-shaped subgraph centered on the matched passage. (b) A semantic query over Vietnamese text retrieves multiple thematically related passages, forming distinct clusters connected through shared linguistic and conceptual nodes.}}
    \label{fig:search_results}
\end{figure}

\subsection{Distribution and Layer Analysis}
\label{sec:dis_layer_analysis}

Initial accuracy tests showed strong linguistic results, with segmentation achieving approximately ${90\%}$ accuracy in annotated samples. Entity recognition was strong for names and places but weaker for abstract terms. Structural analysis confirms the research's focus on deep linguistic and semantic modeling (Figures  \ref{fig:linguistic_density}, \ref{fig:semantic_network}, and \ref{fig:search_results}). The KG exhibits a clear emphasis on linguistic and textual information, with nodes related to these forms constituting nearly ${70\%}$ of all nodes, and the significant presence of \text{EMBEDDING} nodes (${13.9\%}$) further highlighting its reliance on a semantic layer for advanced information retrieval. This linguistic foundation is reinforced by the edge distribution, where the \text{APPEARS\_IN} relationship dominates (${41.3\%}$), confirming robust connections within the textual structure, while the crucial inclusion of \text{HAS\_SEMANTIC\_REPRESENTATION} and \text{BELONGS\_TO\_CLUSTER} relationships among the top relations validates the strong connectivity between textual units and their abstract semantic meanings, enabling sophisticated understanding and organization of information.

\comment{
\subsection{Error Analysis}\label{sec:error_analysis}
From the validation in Section~\ref{sec:dis_layer_analysis}, remaining 10\% of errors typically stem from:
\begin{itemize}
    \item Complex intertextuality: Passages where the expert's commentary and ancient quotes are deeply interwoven without explicit markers, occasionally causing the semantic coherence score to trigger a false boundary.
    \item Phonetic ambiguity (e.g., \cjkterm{樂}; see Section~\ref{sec:linguistic_processing}) cannot be resolved by expanding the contextual window, implying that error reduction in this category requires architectural rather than parametric changes.
\end{itemize}
These cases are addressed through cross-layer connections: the Commentary Layer provides supplementary context that can disambiguate the Linguistic Layer when embedding-based methods fail, effectively distributing the interpretive burden across the graph rather than concentrating it in a single processing step.
}

\begin{figure}[t!]
    \centering
    \includegraphics[trim={0 0 0 0},clip, width=0.6\columnwidth]{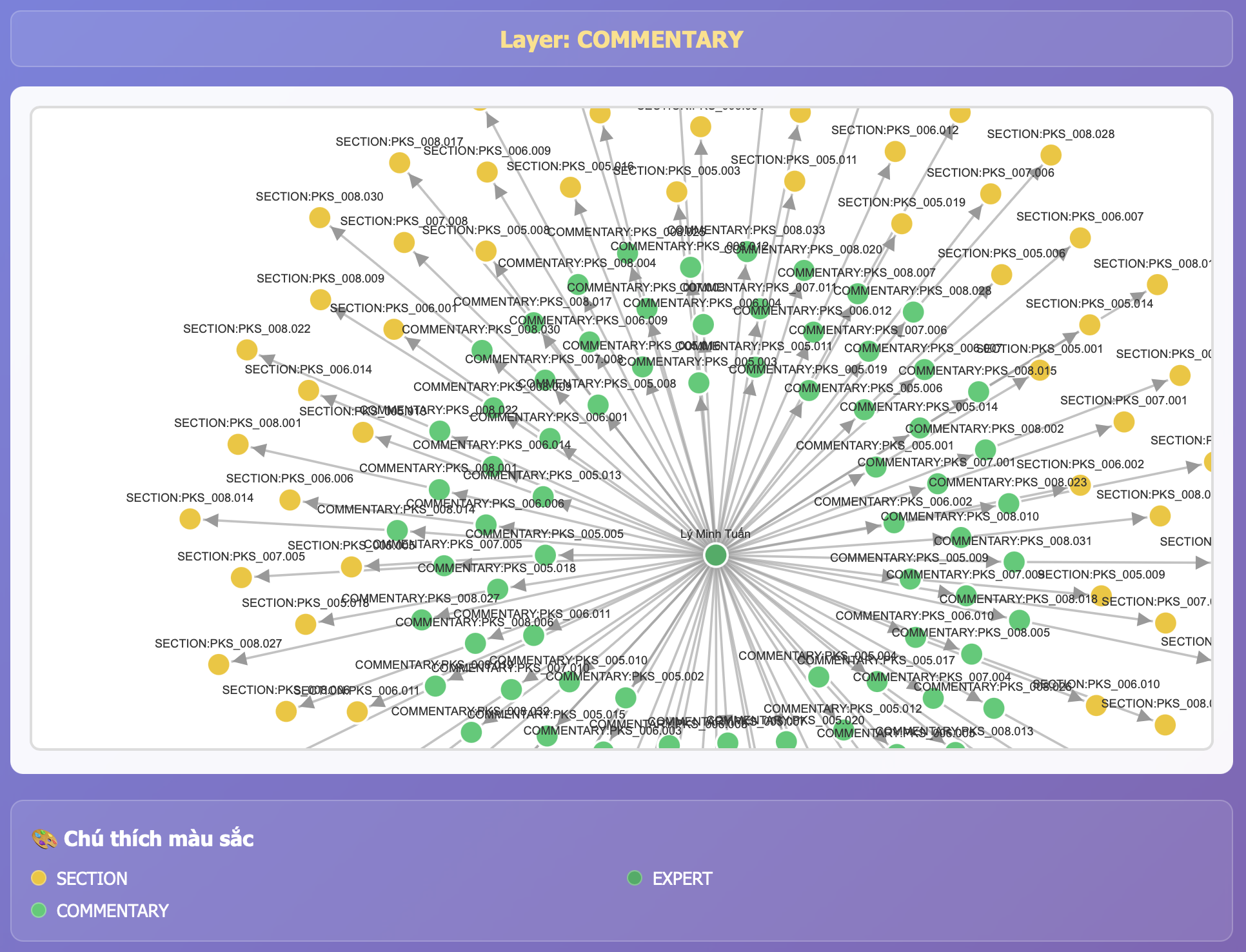}
        \caption{Commentary Layer structure. The central EXPERT node (Ly Minh Tuan) connects directly to COMMENTARY nodes, illustrating the highly centralized and dense structure of the interpretive layer (Density: 0.004149).}
        \label{fig:commentary_layer}
\end{figure}

\subsection{Density Interpretation and Cross-Layer Connectivity}

\begin{figure}[t!]
    \centering
    \includegraphics[trim={0 0 0 0},clip, width=0.6\columnwidth]{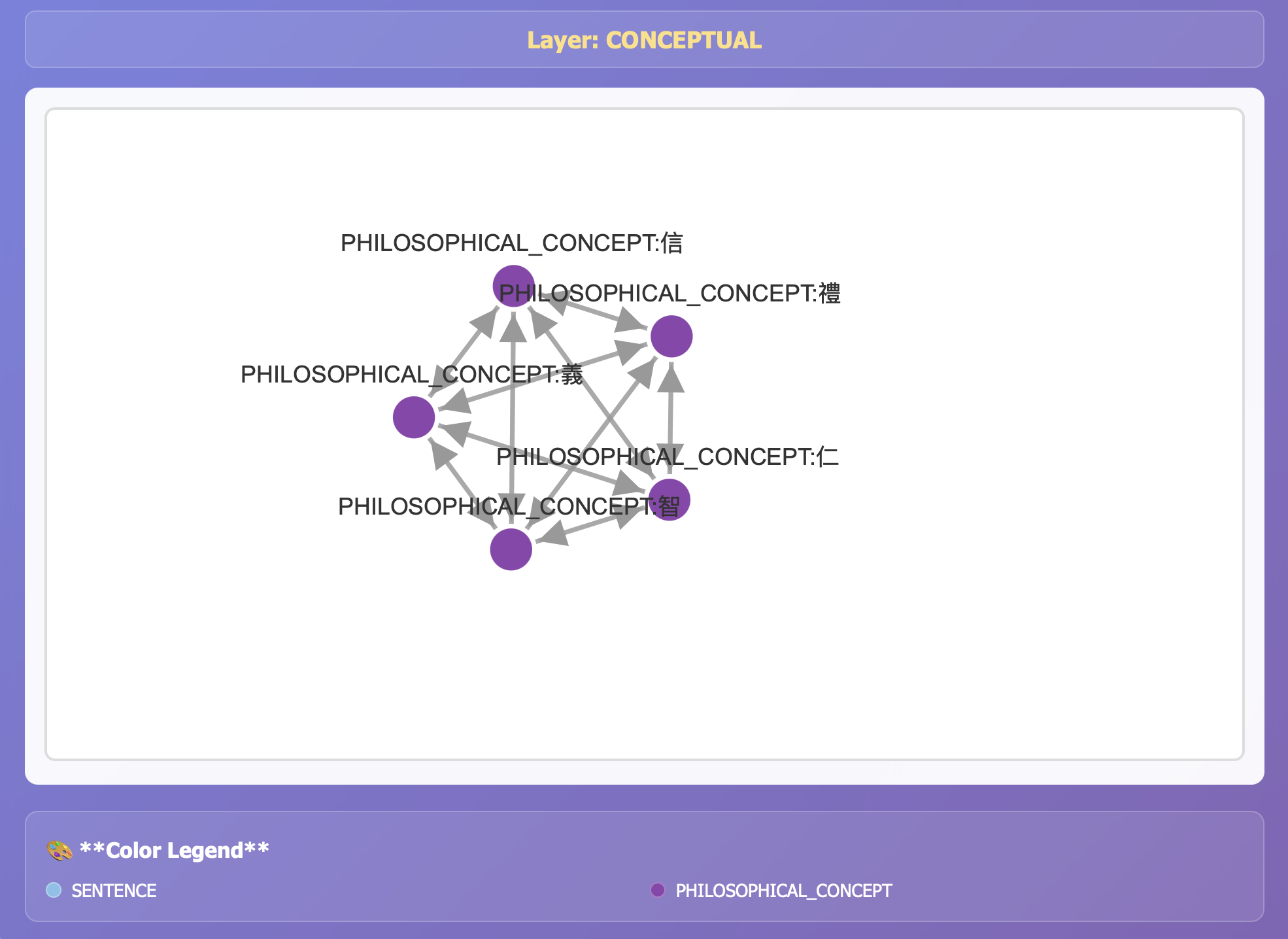}
    \caption{Structural analysis of Conceptual Layer (Density: $0.001368$). The visualization demonstrates that the Philosophical Concept nodes, representing core virtues (\cjkterm{仁}, \cjkterm{義}, \cjkterm{禮}, \cjkterm{智}, \cjkterm{信}), form tightly integrated clusters. This high degree of co-occurrence and relational density within the Conceptual Layer is structurally sound, indicating that these philosophical concepts form integrated networks essential for effective multi-hop reasoning and comparative analysis.}
        \label{fig:conceptual_layer}
\end{figure}

The comparison of individual layer densities offers crucial insights into the graph's architecture: the significantly higher density observed in the Commentary (density: ${0.004149}$) and Conceptual (density: ${0.001368}$) layers (Figures \ref{fig:commentary_layer} and \ref{fig:conceptual_layer}) is structurally sound, indicating that scholarly notes and philosophical concepts form tighter, more integrated clusters essential for effective multi-hop reasoning. Furthermore, the ${12.6\%}$ of total edges dedicated to cross-layer connections serve as a critical bridge, facilitating sophisticated, multi-hop queries that traverse distinct information domains, such as linking a specific sentence to a broader philosophical concept or an identified speaker, thereby validating the complex structural design outlined in the methodology.

\begin{figure}[t!]
    \centering
    \includegraphics[width=0.6\columnwidth]{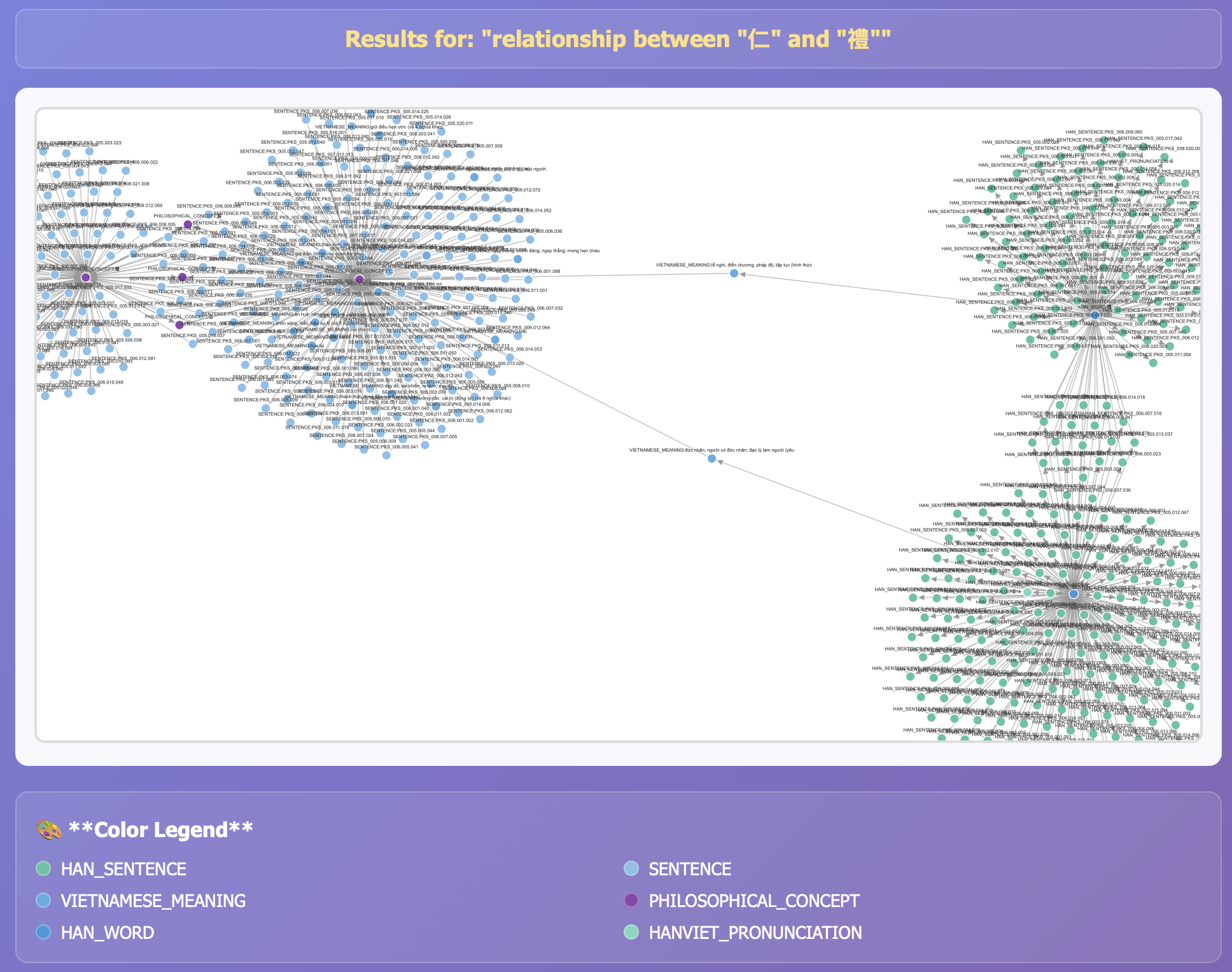}
    \caption{Complex concept query result. The result for "relationship between \cjkterm{仁} and \cjkterm{禮}" illustrates the initial subgraph expansion, showing two large clusters connected by Linguistic and Semantic nodes.}
    \label{fig:query_complex_concept}
\end{figure}

\subsection{Conceptual Tracing and Intertextuality}

Within this KG, entities such as Confucius, Mencius, and their disciples were meticulously extracted, alongside key concepts like \cjkterm{仁} (benevolence) and \cjkterm{禮} (ritual propriety) (Figure~\ref{fig:query_complex_concept}). The application of embedding-based similarity proved instrumental in tracing complex overarching themes, including governance, moral cultivation, and education, thereby moving retrieval capabilities beyond mere keyword matching towards genuine thematic discovery. For instance, these embeddings revealed clear and significant parallels between passages found in The Mencius and The Analects concerning the principles of benevolent leadership. Furthermore, the constructed networks effectively unveiled intricate relationships such as teacher-disciple links, nuanced commentary interpretations, and direct quotations that span across different books within the collection.

\section{\comment{Preliminary User Evaluation}} 
We conducted a pilot study to explore the pedagogical potential and scholarly utility of the system. 
\subsection{\comment{Study Protocol and Participant Recruitment}}
\comment{
We recruited six participants (three males, three females, aged from 22 to 30) from a local university via emails and snowball sampling. The participants were graduate students, specialized in Educational Science and had prior academic exposure to Philosophy, allowing them to provide informed and critical feedback on both the instructional design and the system's scholarly depth. 
At their arrival at the experiment environment, the participants were asked to fill a consent form. 
Then each participant used the system to complete a set of targeted analytical tasks under ethical guidelines of the hosting institution.
The set of tasks the participants had to complete includes:
\begin{itemize}
    \item \textbf{Concept Tracing:} Tracking the evolution of Ren (\cjkterm{仁} - Benevolence) from The Analects to The Works of Mencius.
    \item \textbf{Comparative Analysis:} Comparing how different speakers discuss the virtue of Li (\cjkterm{禮} - Ritual/Propriety) using the Speaker and Semantic layers.
    \item \textbf{Intertextual Exploration: } Utilizing "Exact Text Search" to locate specific commentary nodes bridging multiple textual segments.
\end{itemize}
}
\comment{
After completing the tasks, we conducted semi-structured post-study interviews to elicit participants’ qualitative perceptions, experiences, and feedback regarding the system. The interviews primarily focused on three aspects: learning benefits (Q1), the helpfulness of the AI components (Q2), and overall system usability (Q3). 
All interviews were audio-recorded to support subsequent transcription and analysis.}
\subsection{\comment{Knowledge Integrity and Interpretive Multiplicity}}
\comment{A core challenge in modeling classical texts is the ambiguity of commentaries that often apply to multiple sentences. To ensure scholarly integrity, Graphilosophy avoids arbitrary selection; instead, it implements multi-directional linking by creating concurrent edges for a single node. This decision ensures interpretive multiplicity, allowing users to cross-examine different meanings across layers. While this increases informational density and may cause "visual clutter," it is a calculated trade-off to prioritize data transparency and preserve the original text's complexity over oversimplification.}

Based on this design decision, concept maps, faceted search, and passage-to-concept explanations were successfully deployed in classroom settings for the study, enabling learners to trace philosophical concepts and explore intertextual relationships more effectively (Figure~\ref{fig:query_complex_concept}). The case study confirmed that computational tools can clarify conceptual structures and highlight connections across The Four Books.
\subsection{\comment{Results and Feedback}}

\comment{On average, each participant spent approximately 60 minutes in the user study, including 45 minutes completing the tasks and 15 minutes participating in the post-study interview. The audio recordings of the interviews were transcribed verbatim and analyzed using an inductive thematic analysis approach to identify major themes related to system utility and usability. Initially, one researcher developed the thematic codes through iterative, data-driven coding of the transcripts. These codes and the resulting themes were subsequently reviewed and discussed with two additional researchers to resolve discrepancies and reach consensus.}
\comment{
\paragraph{User Response Summary}
Participant responses were characterized by the frequency of positive sentiment versus reported concerns (Table~\ref{tab:user_feedback}). The qualitative data gathered from the user study was analyzed and categorized into three primary themes: Scholarly Utility, Visual Complexity and AI Integration Pedagogical Support as below.
}
\begin{table}[t!]
\comment{
\centering
\caption{\comment{Summary of participant response trends ($n=6$).}}
\label{tab:user_feedback}
\begin{tabular}{lcl}
\toprule
\textbf{Evaluation Dimension} & \textbf{Positive} & \textbf{Primary Concerns / Suggestions} \\ \midrule
Learning Benefit (Q1) & 4/6 & Visual clutter in complex subgraphs \\ 
AI Helpfulness (Q2) & 5/6 & Text formatting and lack of visual links \\ 
System Usability (Q3) & 3/6 & Language mix and navigation flow \\ \bottomrule
\end{tabular}
}
\end{table}

\comment{
\begin{itemize}
    \item \textbf{Scholarly Utility:} A majority of participants noted that the graph-based approach made intertextual relationships visible and manageable. One user observed: "The graph-based approach makes intertextual relationships visible and manageable". The participants generally confirmed the value of concept tracing and relation visualization, noting that the system improved comprehension and supported course assignments.
    \item \textbf{Visual Complexity and Interaction Flow of Concept Graph:} Users reported that the interface became "messy" or "overwhelming" when dense layers were active (i.e., when many nodes and edges were displayed at once). This confirms the trade-off; while multi-directional linking preserves interpretive richness, it requires better filtering for non-experts. While many praised the aesthetics and innovative design of the interface, others recommended a more consistent flow for exploration of the interface. Some suggested simplifying the visualization by showing only the most relevant nodes, enlarging or reformatting chatbot text for readability, and integrating in-app guidance to help new users navigate the system. 
    \item \textbf{AI Integration for Pedagogical Support:} The Gemini integration was lauded for accuracy and speed. Students highlighted that AI responses from the integrated Gemini chatbot were often clear and usefulHowever, it was criticized for a lack of "animation or clear visual cues" linking chatbot explanations to graph nodes, and for verbose, unformatted text blocks.
\end{itemize}
}


\paragraph{Study Conclusion}

The preliminary study suggests that Graphilosophy improves accessibility to The Four Books while strengthening scholarly rigor through transparent representation and clear provenance of interpretations. The multi-layer graph structure and AI-assisted interaction show strong potential for both research and pedagogical use, particularly in clarifying conceptual organization and intertextual relationships. 

At the same time, the study highlights areas for improvement, including clearer labeling, reduced visual density, and refinement of chatbot output. Limitations remain in segmentation accuracy, the handling of philosophically ambiguous terminology, and the capacity of semantic embeddings to capture nuanced meaning. 

Furthermore, as a small exploratory pilot  with a small sample size ($n=6$), the findings also call for larger and more diverse studies to assess robustness, generalizability, and long-term educational impact.

\section{Discussion}
\comment{
\subsection{Societal and Ethical Dimensions}
\label{sec:ethical}
\subsubsection{Interpretive Authority and Source Selection}
\label{sec:interpretive_authority}
The selection of Ly Minh Tuan's commentary as the primary interpretive lens reflects deliberate pedagogical priorities: bilingual accessibility for 
Vietnamese learners and educational orientation over purely academic discourse. 
However, this choice inherently privileges a Vietnamese interpretive tradition, potentially marginalizing Korean scholarly interpretations and Japanese readings. 
\comment{Importantly, Ly Minh Tuan's work already incorporates Zhu Xi's \textit{Sishu Jizhu} as a foundational reference, translating and annotating 
it alongside his own commentary; the system therefore mediates this neo-Confucian Chinese tradition indirectly rather than excluding it entirely.} The framework's 
horizontal scalability directly addresses the remaining gap: the ontology supports multiple expert nodes, allowing future work to incorporate Korean Samaejip commentaries without restructuring the core architecture. 
\subsubsection{Translation as Cultural Politics}
Translations are never neutral \citep{klein2020}. Classical concepts resist
perfect mapping to modern Vietnamese:
\begin{itemize}
    \item \cjkterm{仁} (Ren): Translated as ``\textit{Nhân}'' (benevolence), but the 
    Vietnamese term carries Buddhist connotations absent in the original Confucian 
    usage. Ly Minh Tuan addresses this by expanding \cjkterm{仁} through commentary nodes 
    that explain it as ``the totality of all virtues'' rather than a single 
    moral quality.
    \item \cjkterm{禮} (Li): Rendered as ``\textit{Lễ}'' (ritual/propriety), but the 
    Vietnamese term emphasizes ceremonial aspects while the Classical Chinese 
    encompasses broader social norms. The system represents this semantic gap 
    through translation relations that preserve multiple meaning 
    nodes rather than collapsing to a single translation.
\end{itemize}
\subsubsection{Algorithmic Mediation and Possible Distortion}
Graph structures prefer discrete, named relations and may struggle with ambiguity, 
indirect allusions, or deliberate contradictions characteristic of philosophical
texts~\citep{drucker2017}. For example, the \textit{Analects} presents seemingly contradictory statements about \cjkterm{仁} that resist single-relation encoding. Our system 
addresses this through:
\begin{enumerate}
    \item \textbf{Multi-hop queries:} Allowing users to traverse multiple relation 
    paths rather than expecting single-edge answers.    
    \item \textbf{Commentary integration:} Expert annotations provide interpretive context that disambiguates apparent contradictions.    
    \item \textbf{Semantic clustering:} The similarity relation groups thematically related passages regardless of surface-level contradiction, enabling users to explore conceptual tensions.
\end{enumerate}
}

\begin{figure}[t!]
    \centering
    \includegraphics[width=\columnwidth]{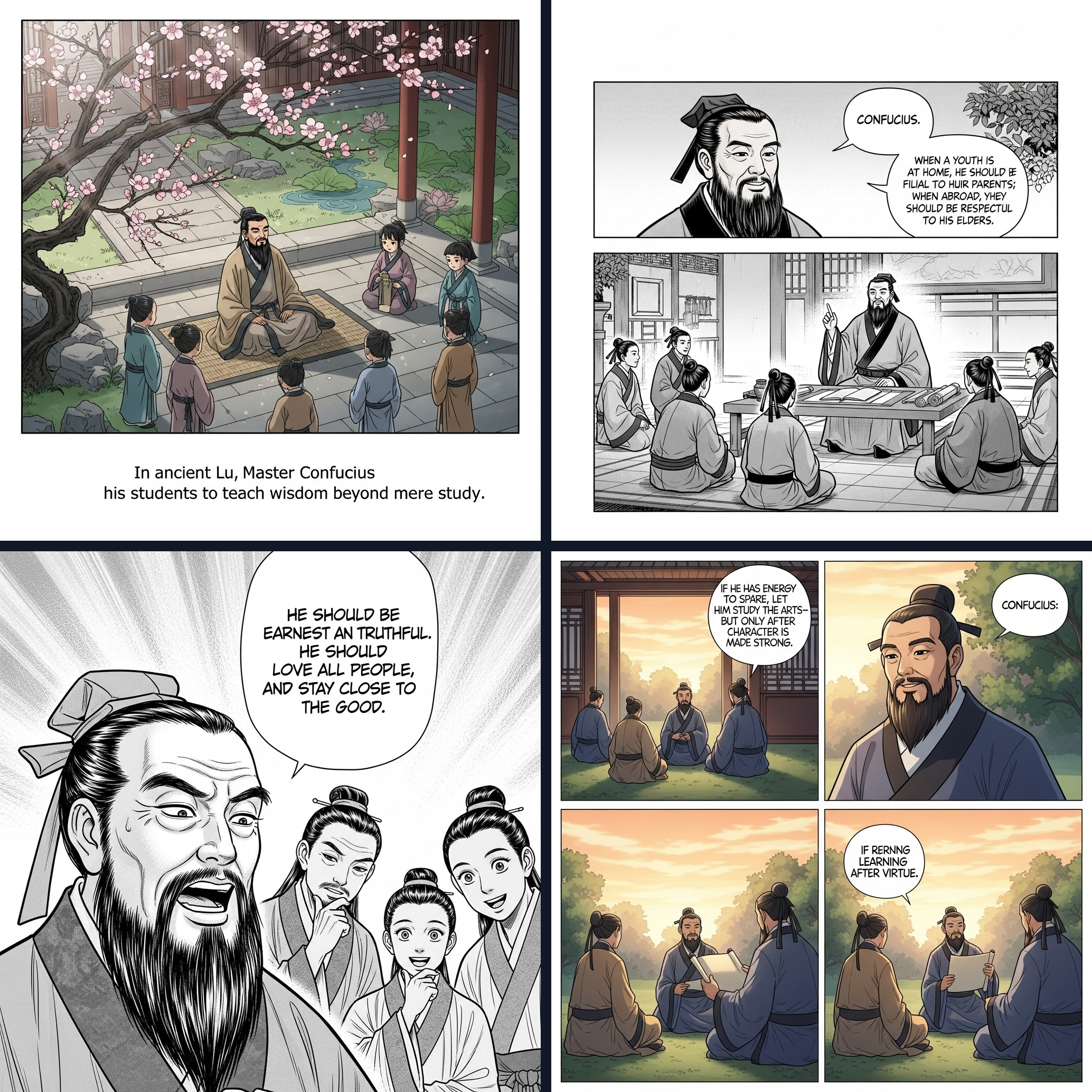}
    \caption{AI-generated narrative visualization from The Analects.}
    \label{fig:comic_page}
\end{figure}

\begin{figure}[t!]
\centering
\subfloat[Liang's cluttered study. Scrolls lie open everywhere, ink stains the desk, an unstrung bow rests among account books.]{\includegraphics[width=0.24\textwidth]{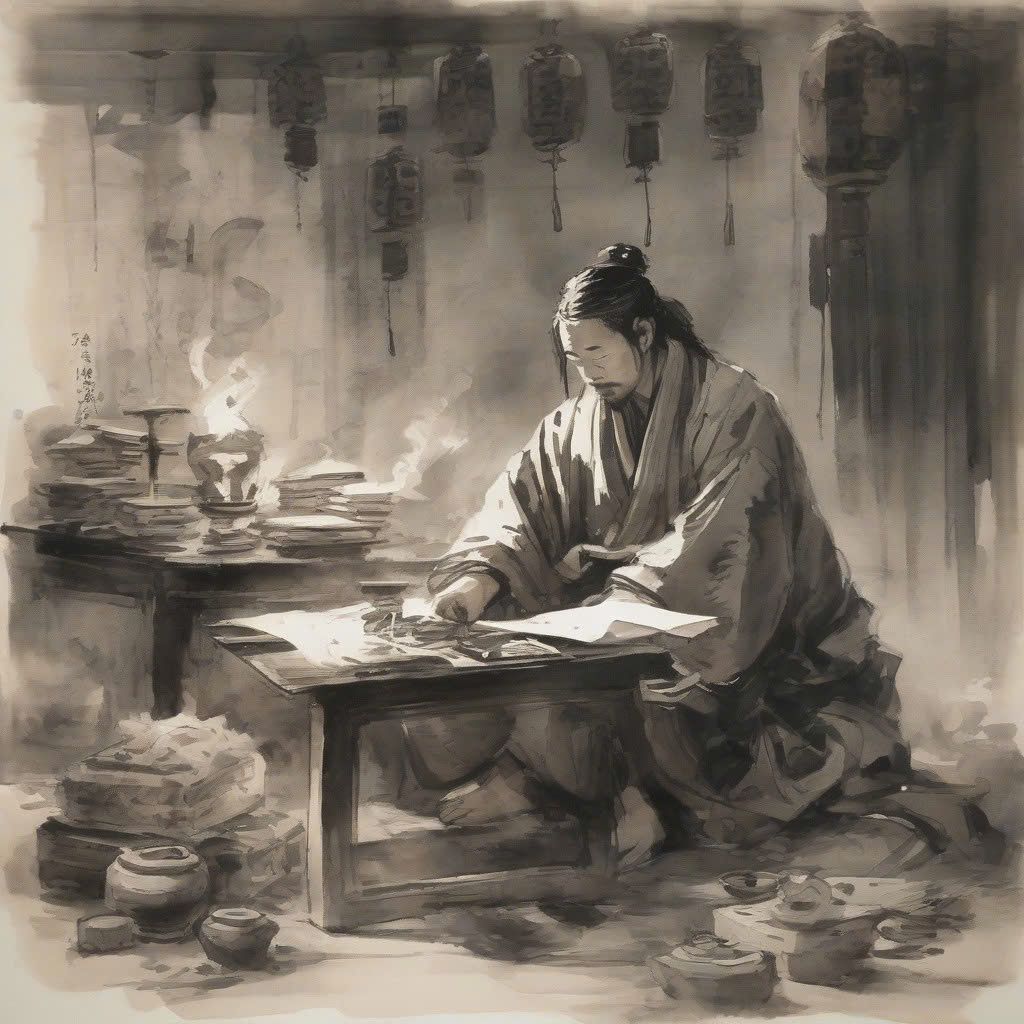}}
\hfill
\subfloat[A tranquil garden. Master Chen sits beneath willow, sipping tea in stillness.]{\includegraphics[width=0.24\textwidth]{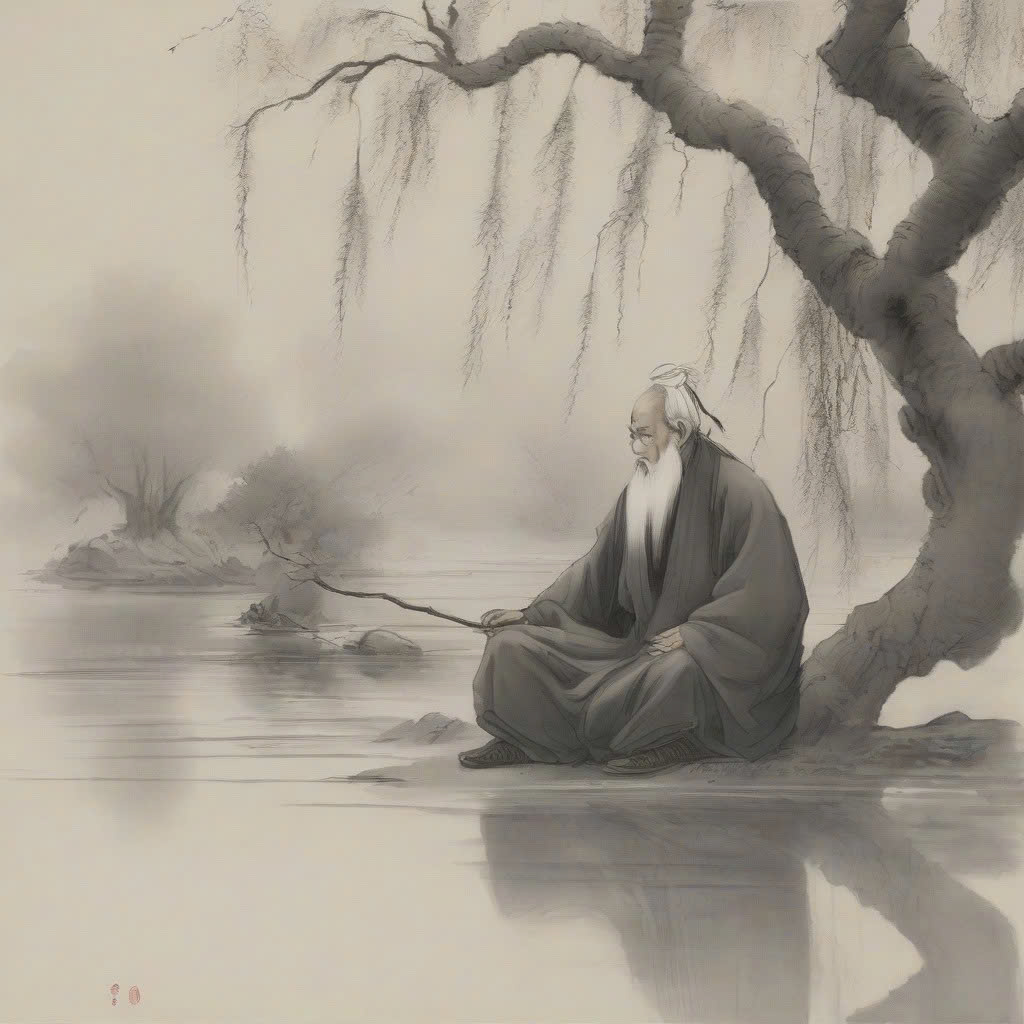}}
\hfill
\subfloat[A small stone table holding a nearly dead bonsai with dry soil and withered branches.]{\includegraphics[width=0.24\textwidth]{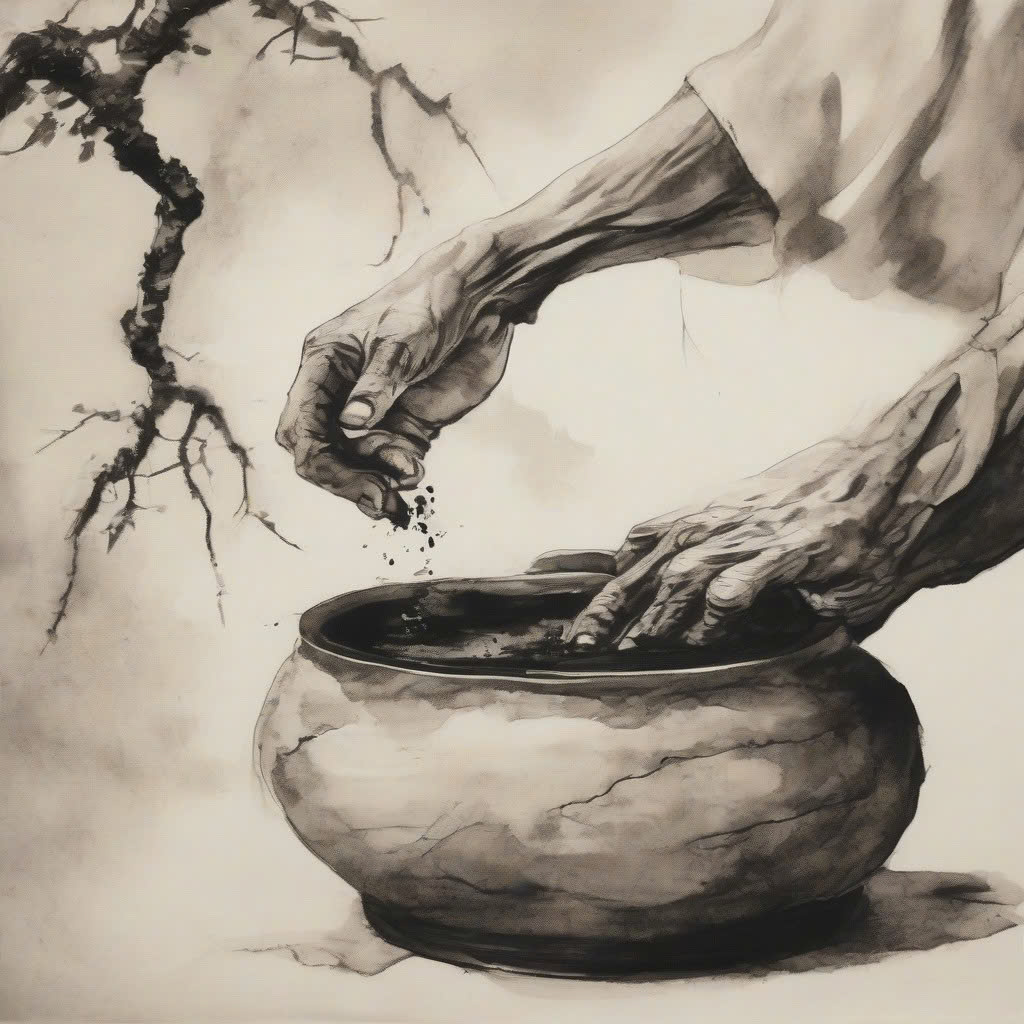}}
\hfill
\subfloat[Liang works anxiously in the garden. Soil splashes, branches fall in uneven cuts, water floods the pot.]{\includegraphics[width=0.24\textwidth]{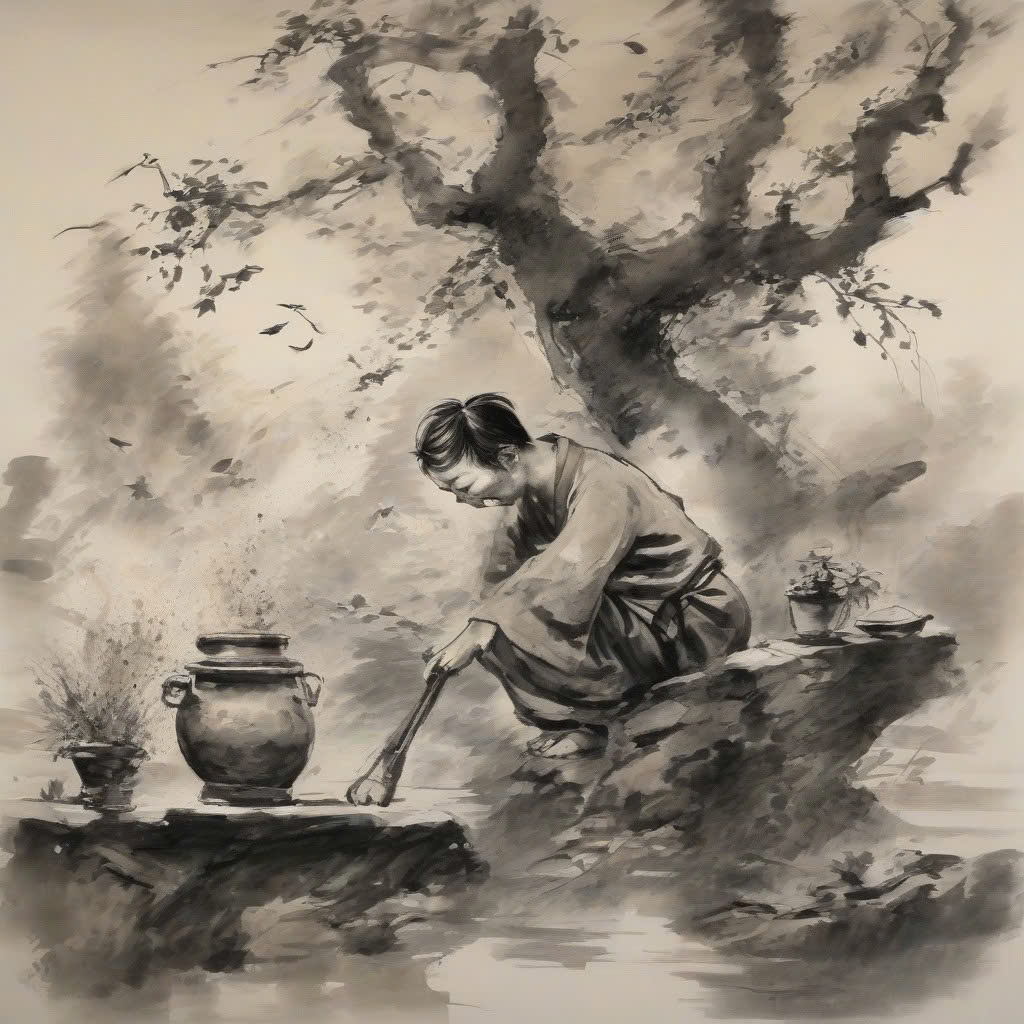}}\\[2pt]
\subfloat[Evening in the garden. The bonsai looks worse—soil soggy, leaves shriveled further.]{\includegraphics[width=0.24\textwidth]{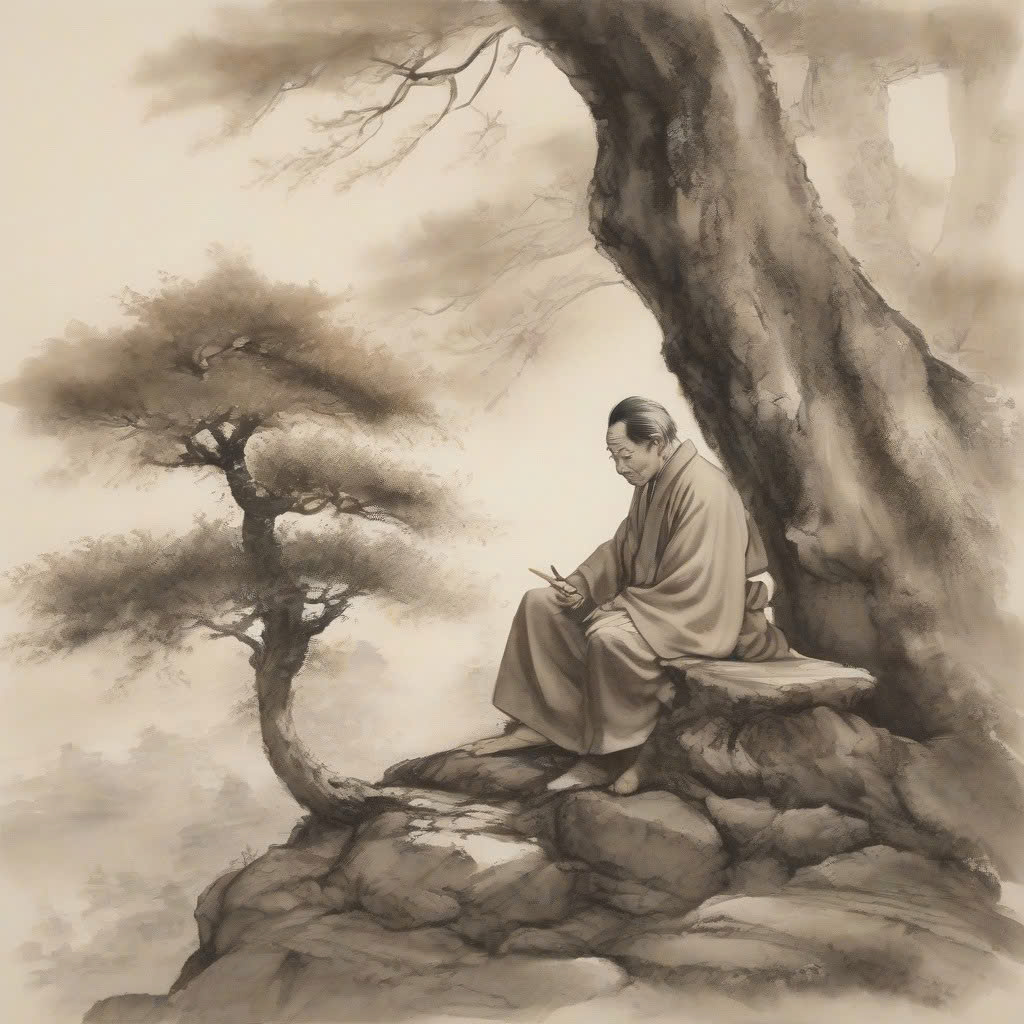}}
\hfill
\subfloat[Morning light. Liang gently loosens roots, replaces soil, waters lightly, and places the bonsai in soft sunlight.]{\includegraphics[width=0.24\textwidth]{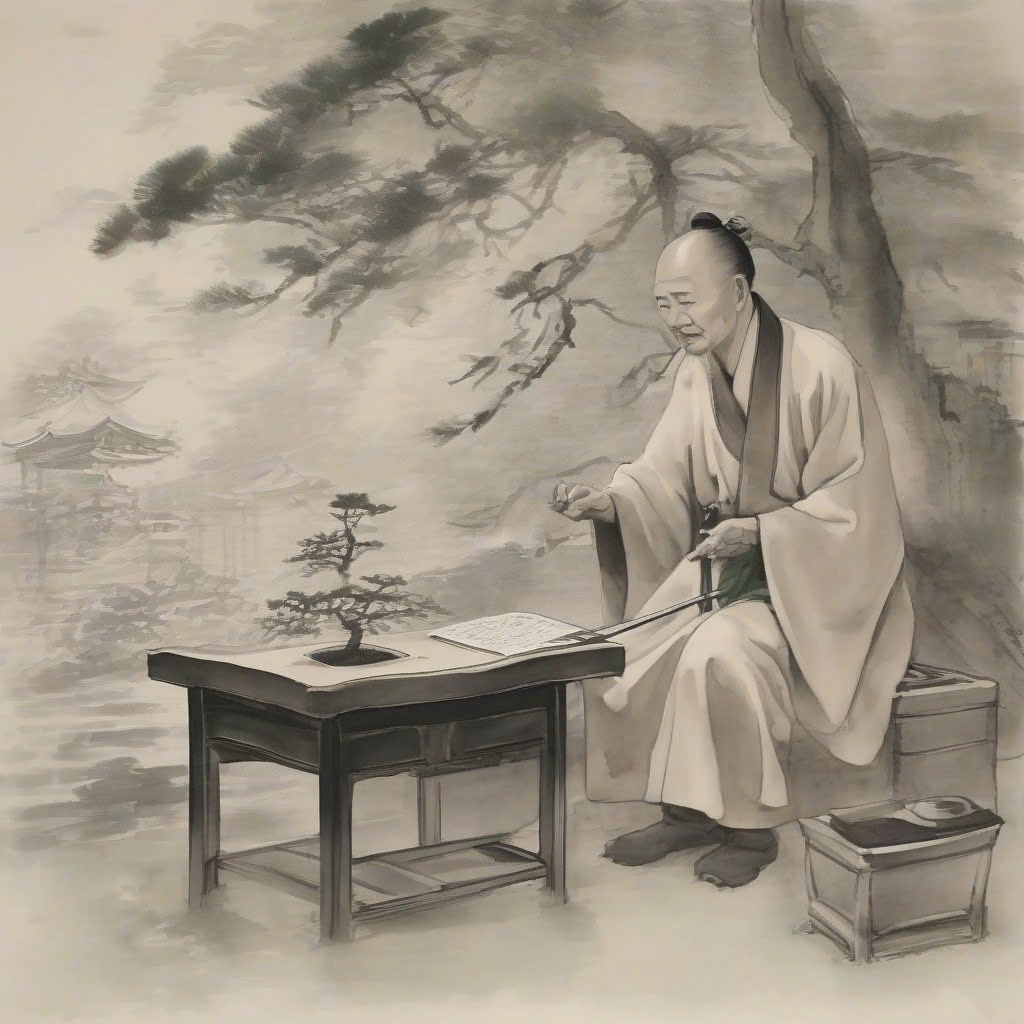}}
\hfill
\subfloat[New shoots sprout from the bonsai's branches. Morning dew glimmers on young leaves.]{\includegraphics[width=0.24\textwidth]{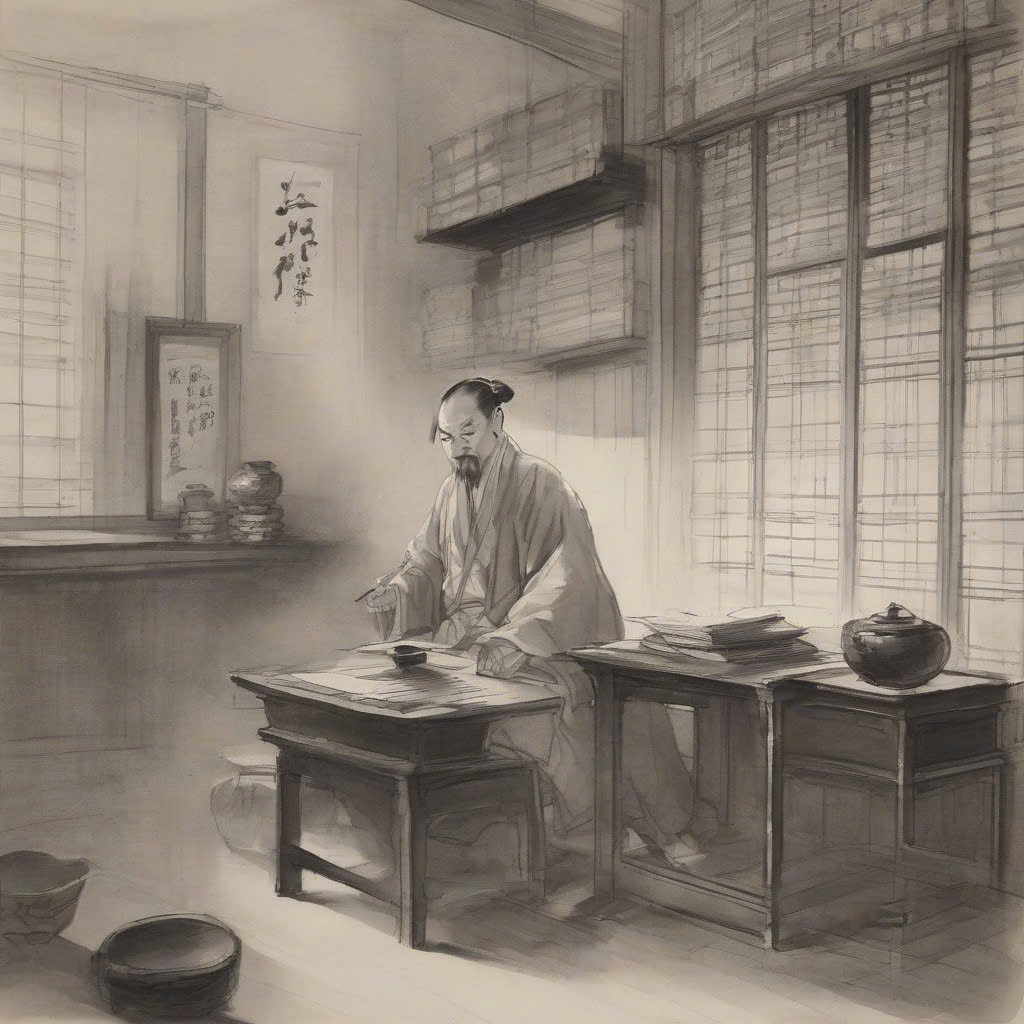}}
\hfill
\subfloat[Liang's study is tidy. A single scroll is open, brush ready, bow and account books neatly stored away.]{\includegraphics[width=0.24\textwidth]{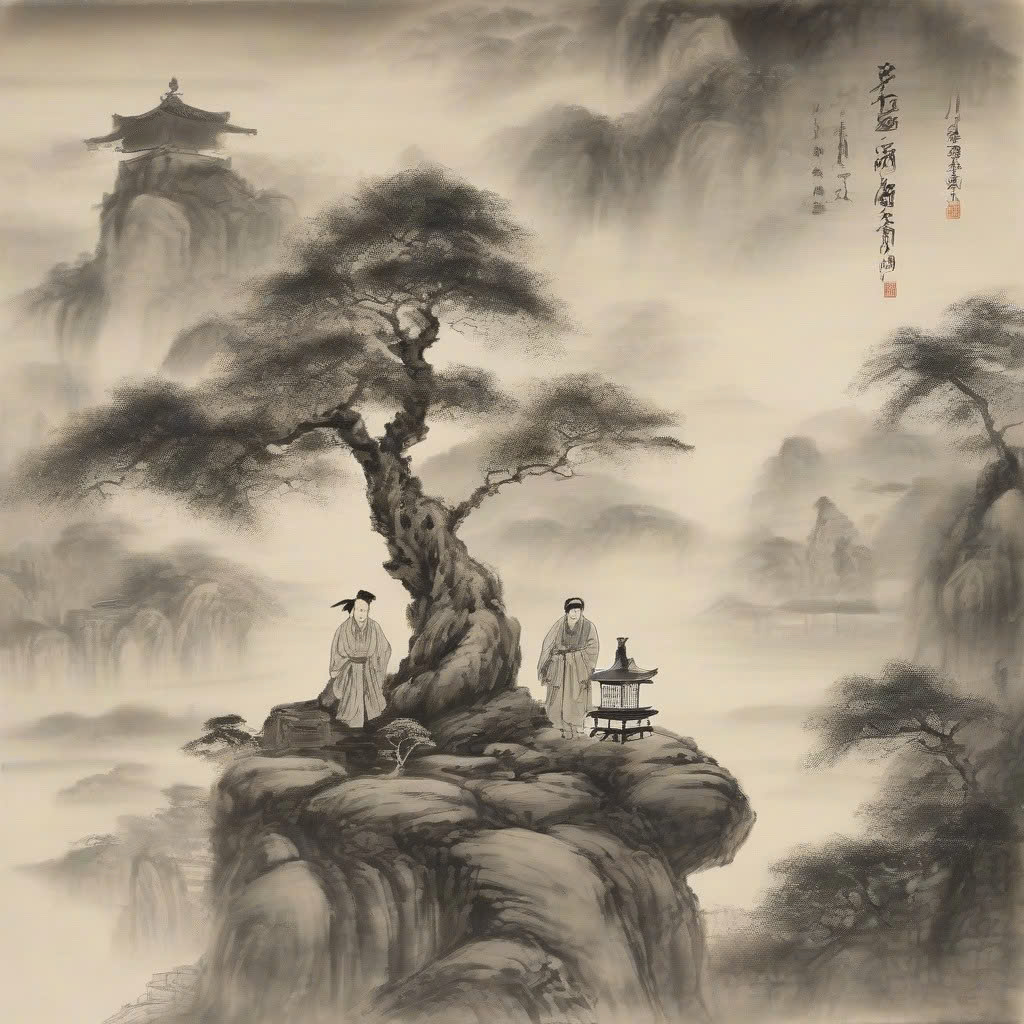}}
\caption{\comment{Qualitative results of the Philosophy-Unfolded – The Great Learning (\cjkterm{大学} / Đại Học) system.  Eight visual narratives depict sequential moral progressions derived from the canonical text and commentary through multimodal generative interpretation.}}
\label{fig:a4_results}
\end{figure}

\subsection{Pedagogical Applications}

Building on the pilot study, we extend the multi-layer KG into generative storytelling by transforming philosophical passages into sequential visual narratives. Figures~\ref{fig:comic_page} and \ref{fig:a4_results} illustrate how classical commentary bridges interpretive scholarship and AI-driven creativity. This prototype functions as a collaborative co-creation platform, assisting users with narrative composition, panel organization, and visual coherence. Ultimately, this framework lays the groundwork for AI-assisted cultural heritage storytelling, merging KGs, generative models, and interactive design to reimagine and preserve Confucian philosophy in accessible, multimodal formats. 

\subsection{Limitations}
\label{sec:limitations}
While Graphilosophy demonstrates the potential of integrating NLP and knowledge 
graphs for classical text analysis, several limitations warrant acknowledgment.

\paragraph{Sample Size and Generalizability}
The preliminary user evaluation involved only six participants from a single institution, limiting the generalizability of user feedback. All participants were graduate students in Educational Science with prior exposure to philosophy, which may not reflect the broader target audience of non-expert learners. Future work should include larger, more diverse cohorts across multiple educational contexts and cultural backgrounds to validate the system's pedagogical effectiveness.


The scope of interpretive authority is further constrained by the dataset's 
reliance on a single scholarly edition; the implications of this choice, and 
the architectural provisions for expanding it, are discussed in 
Section~\ref{sec:interpretive_authority}.

\paragraph{Disambiguation Accuracy}
The system achieves varying accuracy across linguistic challenges, including homophone disambiguation (\cjkterm{人} vs.\ \cjkterm{仁}), phonetic ambiguity (\cjkterm{樂} as Lạc/Nhạc), and speaker attribution. Characters with sparse contextual cues or multiple valid readings in philosophical contexts remain challenging for embedding-based disambiguation. 


This limitation is most acute for characters whose ambiguity is phonetic rather than semantic, a distinction elaborated in Sections~\ref{sec:linguistic_processing} and~\ref{sec:error_analysis}.

\paragraph{Implicit Linguistic Features}
\comment{Beyond phonetic ambiguity, the system does not currently attempt ellipsis recovery or anaphora resolution, a structural limitation of Classical Chinese that affects speaker-attribution queries more broadly (see Section~\ref{sec:error_analysis} for concrete instances). Addressing this would require syntactic augmentation beyond the embedding-based approach adopted here.}

\paragraph{Evaluation Methodology}

\begin{table}[!t]
\comment{
\centering
\caption{\comment{Retrieval performance comparison between BM25, Semantic, and Hybrid approach.}}
\label{tab:retrieval-results}
\begin{tabular}{@{}lcccc@{}}
\toprule
\textbf{Metric} & \textbf{BM25} & \textbf{Semantic (E5)} & \textbf{Hybrid} & \textbf{Best Performer} \\ \midrule
P@1             & 0.773         & \textbf{1.000}         & \textbf{1.000}  & Hybrid/Semantic  \\
P@3             & 0.652         & \textbf{1.000}         & \textbf{1.000}  & Hybrid/Semantic \\
P@5             & 0.564         & \textbf{1.000}         & \textbf{1.000}  & Hybrid/Semantic\\
P@10            & 0.505         & \textbf{1.000}         & 0.995           & Semantic\\
MRR             & 0.773         & \textbf{1.000}         & \textbf{1.000}  & Hybrid/Semantic\\
NDCG@5          & 0.770         & \textbf{1.000}         & \textbf{1.000}  & Hybrid/Semantic\\
NDCG@10         & 0.766         & \textbf{1.000}         & 1.000           & Semantic\\ \bottomrule
\end{tabular}
}
\end{table}

The retrieval evaluation (Table~\ref{tab:retrieval-results}) used a synthetic test corpus containing Confucian concepts and unrelated modern topics as true negatives. While this demonstrates discriminative power, it does not reflect the nuanced relevance judgments required for philosophical inquiry. More rigorous evaluation with expert-annotated relevance judgments from domain scholars would strengthen validity claims. Additionally, the perfect precision scores (P@1 = 1.0) may reflect the controlled nature of the test set rather than real-world retrieval performance.
\paragraph{Scalability Validation}
Although the framework is designed for horizontal and vertical scalability, we have not yet validated performance with substantially larger corpora. The current graph of 16,468 nodes and 71,249 edges represents a single commentary tradition on four books. Extending to multiple commentary traditions or additional philosophical texts may introduce computational and ontological challenges not yet encountered.
\paragraph{Visual Complexity Trade-off}
User feedback consistently identified visual clutter as a usability concern, particularly when dense layers are active. The current implementation 
prioritizes interpretive multiplicity over visual simplicity a deliberate design choice that may limit accessibility for non-expert users. Future iterations should explore progressive disclosure mechanisms or adaptive filtering to balance scholarly completeness with user-friendly visualization.

\section{Conclusion}

This paper demonstrates how NLP and KG construction can deepen engagement with classical philosophical corpora in  \textit{The Four Books}, a domain where linguistic concision, interpretive plurality, and centuries of commentary tradition have resisted stable computational representation. By curating a trilingual corpus, designing a six-layer ontology, and anchoring retrieval in semantic embeddings, Graphilosophy advances concept tracing, 
intertextual analysis, and AI-assisted pedagogy while keeping interpretive authority visible and contestable.

\comment{The two research questions framing this study asked whether a multi-layered KG can represent interpretive plurality without reducing it, and whether such representation can support learning without foreclosing the openness of the original texts. The evidence presented here suggests that both questions admit qualified affirmative answers, and that the qualification in each case points 
toward the same underlying tension. Graphilosophy's architecture preserves translational asymmetry, distributes commentary authority, and encodes polysemy as a navigable feature rather than an error to be corrected; yet the very density that enables this fidelity is precisely what makes the system cognitively demanding for non-expert users. This friction is not an implementation failure but a structural one: fidelity to philosophical complexity and accessibility for broad audiences are not straightforwardly compatible design goals.}

Our preliminary evaluation indicates that although maintaining interpretive multiplicity leads to greater visual complexity, it substantially enhances transparency and reinforces the academic rigor of digital hermeneutic analysis. 
Future work will focus on expanding evaluation metrics, and extending the framework to additional Confucian and East Asian philosophical texts. \comment{Resolving the tension between interpretive completeness and pedagogical accessibility, through adaptive filtering, progressive disclosure, or differentiated user pathways, remains the central design challenge ahead, and perhaps the central methodological question for AI-mediated engagement with classical philosophical heritage more broadly.}

\comment{
\section*{Acknowledgments}
This research is funded by Vietnam National University - Ho Chi Minh City (VNU-HCM) under Grant Number B2026-18-17.
}



    


\bibliography{sample-base}

\clearpage
\begin{appendices}

\section{Dataset}

\subsection{Data Source}

The dataset is organized into three primary components:

\begin{itemize}
    \item \textbf{Main Text:} This component includes 2,222 sentences representing the fundamental structural and linguistic units of The Four Books. Key indexing fields include \texttt{file\_id} (book identifier), \texttt{sect\_id} (chapter and section identifier), \texttt{page\_id} (page number), and \texttt{sent\_id} (sentence identifier), which together define the hierarchical structure of the Textual Structure Layer. The linguistic data encompass the original Classical Chinese text (C), the Sino-Vietnamese (Han-Viet) phonetic (V), and the modern Vietnamese translation (M), serving as the foundation for the Linguistic and Semantic Layers.
    
    \item \textbf{Dictionary:}  This component contains 5,344 entries of Classical Chinese characters, including the Chinese form, Sino-Vietnamese (Han-Viet) phonetic, and Vietnamese meanings. A key challenge lies in the polysemy of many characters, each may have multiple entries with distinct meanings or pronunciations. To address this, the system incorporates a semantic consolidation mechanism to merge related entries and minimize redundancy. After semantic consolidation to merge related entries and remove duplicates, the refined dictionary comprises 2,788 unique entries (see Section 3.2).
    
    \item \textbf{Expert Analysis:} This section includes 80 entries of expert-level commentary authored by Ly Minh Tuan, offering contextual and interpretive insights that enrich the Commentary Layer. Each commentary entry is algorithmically linked to its corresponding section (\texttt{sect\_id}) in the main text, allowing cross-referencing between doctrinal passages and their scholarly interpretation.

\end{itemize}

\subsection{Construction Pipeline}

The dataset construction process was organized into three major stages: preprocessing, alignment, and structuring the corpus.

{\textbf{Preprocessing.}} All source documents were converted from PDF into structured digital text through semi-automatic extraction. This process involved several key steps:
\begin{itemize}
\item Digitization and normalization: An optical character recognition (OCR) correction and character standardization library (i.e., PaddleOCR\footnote{\url{https://www.paddleocr.ai/}}) was applied to ensure textual consistency, including the removal of headers, footers, and page artifacts.
\item Segmentation: The base text and commentary were separated yet aligned to preserve interpretive relationships. Texts were divided into sentences or discourse units that matched the logical structure of the original works.
\item Structural encoding: Classical Chinese characters in traditional script were standardized and represented in Unicode. Hierarchical structures, including book, chapter, section, and sentence, were encoded in XML/JSON format.
\item Metadata curation: Provenance information such as edition, commentary author, and variant readings was attached at the segment level.
\item Lexical processing: Glossary sections were automatically detected and parsed using regular expressions (regex). Contextual metadata (book and chapter) was assigned to each lexical entry, ensuring accurate semantic anchoring.
\end{itemize}

{\textbf{Alignment Strategies.}} Distinct alignment strategies were employed for different corpus components:
\begin{itemize}
\item Tri-Parallel Corpus: A rule-based heuristic approach was developed to align the Classical Chinese text with its Sino-Vietnamese (Han-Viet) phonetic and modern Vietnamese translation. Explicit textual cues, such as "Translation:" markers and structural delimiters, were utilized to achieve consistent three-way alignment.
\item Lexical Corpus: Lexical items were extracted from dictionary sections through regex-based parsing and forward-fill propagation of contextual metadata (e.g., book and chapter). Post-processing included deduplication, normalization, and removal of incomplete entries to produce a clean, queryable dictionary resource.
\item Exegesis Corpus: Commentary sections were identified through rule-based parsing and text cleaning. Each commentary entry was linked to its corresponding canonical text and translation via structured metadata. This corpus captures interpretive layers that complement the bilingual and lexical resources.
\end{itemize}

\textbf{Manual Refinement.} We employed a two-stage refinement process consisting of heuristic recovery, in which a simple module was implemented to resolve OCR-induced artifacts (e.g., character substitutions, diacritic distortions, and mid-sentence line breaks), followed by an expert manual audit, during which all textual triplets were manually verified by domain researchers to ensure exact synchronization across linguistic layers. This deterministic procedure guarantees a strict 1:1:1 mapping between Classical Chinese, Sino-Vietnamese, and modern Vietnamese sentence nodes, reflected in identical sentence indices across layers, providing a robust foundation for the multi-layered graph.

\textbf{Structuring and Export.} All corpora were encoded into standardized formats (CSV, XML, and Excel) with unique structured identifiers (\texttt{File.Sect.Page.STC}). This design ensures interoperability among the tri-parallel, lexical, and exegesis corpora, forming a coherent foundation for downstream applications such as ontology-guided knowledge graph construction, semantic retrieval, and cross-lingual analysis.

\subsection{Dataset Description}

The finalized dataset consists of 2,222 segmented sentences, 80 commentary annotations, and a refined lexical dictionary comprising 2,788 entries, including 2,562 unique Classical Chinese characters and 23 domain-specific Confucian concept terms. Ontology instantiation yielded a fully structured knowledge graph containing 16,468 nodes and 71,249 edges, systematically representing entities, semantic relations, and hierarchical interconnections across textual, linguistic, and conceptual layers.

\textbf{Tri-Parallel Corpus.} 
The tri-parallel corpus extends the alignment into a three-layer structure comprising: (1) the original Classical Chinese text, (2) the Sino-Vietnamese (Han-Viet) phonetic, and (3) the modern Vietnamese semantic translation. This corpus includes 2,222 aligned triplets derived from \texttt{\textit{Commentaries on The Four Books}}. The tri-parallel design offers several advantages: it minimizes the gap between orthography and semantics, supports simultaneous phonetic and semantic learning, and enables multi-view NLP tasks such as phonetic-aware translation and multistage alignment. Although the rule-based approach ensures high precision in structurally consistent passages, it remains sensitive to irregular formatting and cases where a single Classical Chinese sentence corresponds to multiple Vietnamese interpretations.

\textbf{Lexical Dictionary.} 
The lexical dictionary is a dictionary-level resource extracted from the glossary sections of \texttt{\textit{Commentaries on The Four Books}}. It contains 5,344 entries, each consisting of a unique identifier (ID), Classical Chinese character, Sino-Vietnamese (Han-Viet) phonetic, modern Vietnamese meaning, and source context (book and chapter of occurrence). This lexical corpus serves as a critical bridge between traditional vocabulary and modern computational linguistics. It underpins knowledge graph construction, facilitates semantic disambiguation, and supports the development of educational and language-learning applications focused on classical texts.

\textbf{Exegesis Corpus.} 
The exegesis corpus encompasses 80 commentary annotations of interpretive commentary derived from \texttt{\textit{Commentaries on The Four Books}}. Each commentary passage is linked to its corresponding canonical text and translation through structured metadata, including book, chapter, and section identifiers. This corpus adds interpretive depth to the overall dataset, enriching the tri-parallel and lexical corpora with contextual explanations, philosophical insights, and scholarly interpretation.

\section{Ontology Relation Generation}

Relations are generated through three distinct methods, each with specific validation requirements.

\paragraph{Fully Automatic (Rule-based)}
These relations are generated through deterministic algorithms without human intervention:

\begin{itemize}[noitemsep]
    \item \textbf{Structural:} \texttt{CONTAINS}, \texttt{APPEARS\_IN}, \texttt{FOLLOWS}, \texttt{HAS\_HAN\_FORM}, generated through hierarchical parsing of document structure and tri-parallel alignment.
    \item \textbf{Linguistic:} \texttt{TRANSLATES\_TO}, \texttt{PRONOUNCED\_AS}, dictionary lookup with contextual embedding disambiguation.
    \item \textbf{Speaker:} \texttt{QUOTES}, pattern-based regex detection for attribution markers (e.g., \cjkterm{子曰}, \cjkterm{孟子曰}, \cjkterm{曾子曰}).
    \item \textbf{Semantic:} \texttt{BELONGS\_TO\_CLUSTER}, \texttt{SIMILAR\_TO}, \texttt{HAS\_SEMANTIC\_REPRESENTATION}, computed from Multilingual-E5-Large embeddings with fixed thresholds.
\end{itemize}

\paragraph{Semi-automatic (Embedding + Verification)}
These relations combine algorithmic generation with sampling-based verification:
\begin{itemize}[noitemsep]
    \item \textbf{\texttt{EXPRESSES\_CONCEPT}:} Character pattern matching against predefined taxonomy, validated through embedding clustering to confirm semantic coherence.
    \item \textbf{\texttt{CONTEXTUALIZES}:} Initial candidates generated via semantic similarity (cosine $>$ 0.75), followed by 10\% sampling-based manual verification.
    \item \textbf{\texttt{RELATED\_TO}:} Co-occurrence frequency analysis combined with manual expert review for philosophical validity.
\end{itemize}
\paragraph{Manual}
These relations are defined based on scholarly classification:

\begin{itemize}[noitemsep]
    \item \textbf{\texttt{PROVIDES\_COMMENTARY}:} Expert attribution (current implementation: single expert Ly Minh Tuan).
    \item \textbf{Taxonomic relations:} \texttt{BELONGS\_TO\_SCHOOL}, \texttt{PART\_OF\_DOMAIN}, domain hierarchy predefined.
    \item \textbf{Concept taxonomy:} 23 core Confucian concepts manually categorized by thematic function.
\end{itemize}

\section{Core Confucian Concepts}
Conceptual Layer is anchored by 23 fundamental Confucian concepts, categorized by thematic function to allow for nuanced tracing of philosophical development. Table \ref{tab:confucian-concepts} presents the complete taxonomy. This taxonomy enables the system to automatically map granular linguistic units to abstract philosophical entities through character pattern matching, preserving interpretive multiplicity while maintaining structural integrity across the graph. For example, when the character \cjkterm{仁} appears in a sentence, the system creates an \texttt{EXPRESSES\_CONCEPT} edge linking that sentence to the \texttt{PHILOSOPHICAL\_CONCEPT:\cjkterm{仁}} node, which is further connected to its categorical grouping (Virtue) via \texttt{RELATED\_TO} edges.

\begin{table}[t!]
\centering
\caption{Core Confucian concepts taxonomy.}
\label{tab:confucian-concepts}
\small
\begin{tabular}{@{}cllll@{}}
\toprule
\textbf{Character} & \textbf{English} & \textbf{Vietnamese} & \textbf{Category} \\
\midrule
\multicolumn{4}{l}{\textit{Cardinal Virtues (\cjkterm{五常})}} \\
\cjkterm{仁} & Benevolence & Nhân & Virtue \\
\cjkterm{義} & Righteousness & Nghĩa & Virtue \\
\cjkterm{禮} & Ritual propriety & Lễ & Virtue \\
\cjkterm{智} & Wisdom & Trí & Virtue \\
\cjkterm{信} & Trustworthiness & Tín & Virtue \\
\midrule
\multicolumn{4}{l}{\textit{Self-Cultivation (\cjkterm{修身})}} \\
\cjkterm{德} & Virtue/Power & Đức & Cultivation \\
\cjkterm{誠} & Sincerity & Chân & Cultivation \\
\cjkterm{正} & Correctness & Chính & Cultivation \\
\midrule
\multicolumn{4}{l}{\textit{Cosmological Foundations}} \\
\cjkterm{道} & The Way & Đạo & Foundation \\
\cjkterm{天} & Heaven & Thiên & Foundation \\
\cjkterm{中} & The Mean & Trung & Harmony \\
\cjkterm{和} & Harmony & Hòa & Harmony \\
\midrule
\multicolumn{4}{l}{\textit{Social Relations (\cjkterm{五倫})}} \\
\cjkterm{孝} & Filial piety & Hiếu & Relation \\
\cjkterm{悌} & Fraternal respect & Đệ & Relation \\
\cjkterm{忠} & Loyalty & Trung & Relation \\
\cjkterm{恕} & Forgiveness/Reciprocity & Thứ & Relation \\
\midrule
\multicolumn{4}{l}{\textit{Learning and Education}} \\
\cjkterm{學} & Learning & Học & Learning \\
\cjkterm{教} & Teaching & Giáo dục & Learning \\
\cjkterm{知} & Knowledge & Tri & Learning \\
\midrule
\multicolumn{4}{l}{\textit{Political Order}} \\
\cjkterm{君} & Ruler & Quân & Social \\
\cjkterm{臣} & Minister & Thần & Social \\
\cjkterm{民} & People & Dân & Social \\
\cjkterm{政} & Government & Chính & Social \\
\bottomrule
\end{tabular}
\end{table}

\section{Semantic Chunking}

The first stage of the pipeline focuses on transforming raw classical and modern texts into structured units while preserving their semantic meaning, which is critical given the complexity and polysemy of Classical Chinese. To effectively process long commentary passages, a semantic-aware adaptive chunking module dynamically segments text based on semantic coherence rather than fixed character limits. Each document is first tokenized and then divided into segments with a maximum length of $L=512$ tokens and a look-ahead overlap of $O=100$. Text encoding is performed using the Multilingual-E5-large model, ensuring that each chunk captures a coherent philosophical argument suitable for subsequent Retrieval-Augmented Generation (RAG) tasks. An automated validation check further improves reliability by switching to a simpler fallback strategy when initial chunk quality is low.

The adaptive chunking module employs cosine similarity of Multilingual-E5-Large embeddings to preserve semantic integrity when segmenting long commentary passages. 
Unlike fixed-length chunking that may split coherent arguments mid-sentence, our approach detects natural topic boundaries through embedding-based coherence analysis.
For a sequence of extracted sentences $S = \{s_1, s_2, \ldots, s_n\}$, each sentence is encoded using the Multilingual-E5-Large model with the instruction prefix 
``passage:'' to obtain embedding vectors. The semantic coherence score for sentence $s_i$ is computed as the average cosine similarity with preceding sentences within a sliding window of size $w = 3$:

\begin{equation}
    \text{coherence}(s_i) = \frac{1}{\min(i, w)} \sum_{j=\max(0, i-w)}^{i-1} 
    \cos\left(\mathbf{e}_{s_i}, \mathbf{e}_{s_j}\right).
\end{equation}

A topic boundary is identified when $\text{coherence}(s_i) < \theta$, where the threshold $\theta = 0.3$ was empirically determined to balance granularity with coherence preservation. When a boundary is detected, the current chunk is finalized and a new chunk begins, ensuring that semantically related content remains together.
The chunking process also enforces size constraints: maximum chunk length $L = 512$ tokens (compatible with downstream RAG tasks) and minimum chunk size $M = 256$ characters (preventing overly fragmented segments). A post-hoc quality validation verifies content coverage $\geq 95\%$; if validation fails, the system falls back to simpler fixed-length chunking to ensure robustness.
This mechanism is particularly important for Ly Minh Tuan's commentary, which often develops philosophical arguments across multiple sentences. By respecting semantic boundaries rather than arbitrary character limits, the adaptive chunker preserves the interpretive integrity essential for downstream retrieval and AI-grounded question answering.

For linguistic processing, a custom dictionary processor is employed to load, validate, and consolidate lexical resources across Classical Chinese, Sino-Vietnamese (Han-Viet) phonetic, and modern Vietnamese layers. 
This module resolves character-level polysemy and harmonizes multiple interpretations to prepare data for the Linguistic Layer. Philosophical concepts are pre-tagged according to a defined taxonomy (e.g., virtue, cultivation), while sentence embedding and semantic similarity computation are performed using the Sentence Transformers framework. Entity and concept extraction builds upon predefined taxonomies (e.g., \cjkterm{仁}, \cjkterm{禮}) in conjunction with dictionary-based lookups and pattern-matching rules, ensuring high coverage and precision in identifying core Confucian concepts.

\section{KG Layer-Specific Construction}

The \textbf{Textual Layer} encodes the canonical hierarchy from \texttt{BOOK} to \texttt{SENTENCE} through \texttt{CONTAINS} relations, forming approximately 2,400 nodes and serving as the structural backbone of the graph.

The \textbf{Linguistic Layer} enriches this structure by linking Classical Chinese sentences to their Sino-Vietnamese (Han-Viet) phonetic and modern Vietnamese translations, and by connecting Classical Chinese words to dictionary entries via \texttt{TRANSLATES\_TO} and \texttt{PRONOUNCED\_AS} relations. This layer contains about 11,600 nodes and 37,700 edges, reflecting the complexity of cross-linguistic alignment.

The \textbf{Conceptual Layer} isolates philosophical notions through character-level pattern matching and associates them with sentences using \texttt{EXPRESSES\_CONCEPT} and \texttt{RELATED\_TO} relations, thereby capturing the underlying philosophical ideas conveyed in the text.

The \textbf{Commentary Layer} introduces expert annotations as \texttt{EXPERT} and \texttt{COMMENTARY\_CHUNK} nodes, sequentially ordered through \texttt{FOLLOWS} relations and connected to the base text via \texttt{EXPLAINS} and \texttt{CONTEXTUALIZES} relations, linking interpretive insights to the canonical material.

The \textbf{Speaker Layer} identifies quotation attributions through rule-based detection (e.g., \cjkterm{子曰}, \cjkterm{孟子曰}), assigning statements to \texttt{SPEAKER} nodes connected by \texttt{QUOTES} relations.

Finally, the \textbf{Semantic Layer} introduces an additional dimension of connectivity by leveraging {multilingual-e5-large} embeddings to link each node with semantically related neighbors, organizing them into \texttt{SEMANTIC\_CLUSTER}s based on cosine similarity thresholds.

\section{NLP Component Evaluation}
\label{sec:nlp_eval}

\begin{table}[!t]
\centering
\caption{Retrieval effectiveness between Adaptive Semantic and Fixed Chunking.}
\label{tab:chunking_eval}
\begin{tabular}{lccc}
\hline
\textbf{Method} & \textbf{Recall@5} & \textbf{NDCG@5} & \textbf{Mean Similarity} \\ \hline
Fixed Chunking (Baseline) & 0.315 & 0.281 & 0.861 \\
\textbf{Adaptive Semantic Chunking (Ours)} & \textbf{0.380} & \textbf{0.333} & \textbf{0.863} \\ \hline
\end{tabular}
\end{table}

\subsection{Semantic Alignment and Retrieval Performance}
To validate the Semantic Layer, we conducted a comparative test between our Multilingual-e5-large embedding model and the traditional keyword-based BM25 baseline. We used a synthetic test corpus containing both Confucian concepts (Relevant) and unrelated modern topics (True Negatives) to measure the model's discriminative power.
The quantitative results across multiple standard retrieval metrics are presented in Table 3.
The empirical data demonstrates that the semantic approach achieves perfect precision and ranking scores ($P@1 = 1.000, MRR = 1.000$). While the BM25 baseline performs reasonably well, its accuracy significantly degrades as the retrieval depth ($K$) increases, dropping to a $P@10$ of 0.505. Our Hybrid method successfully maintains the high precision of the semantic model while utilizing RRF for robust ranking. These results demonstrate strong discriminative power under controlled conditions; their interpretation in open-domain settings is discussed in Section 7.

\subsection{Adaptive Semantic Chunking Performance}
For long commentary passages where fixed boundaries are absent, we evaluated our Adaptive Semantic Chunking against a standard Fixed-Length baseline ($L=512, O=0$). We used a test corpus of 200 annotated queries to measure retrieval effectiveness, as shown in Table \ref{tab:chunking_eval}.
The 20.6\% improvement in Recall@5 and 18.5\% increase in NDCG@5 validate that our semantic-aware segmentation successfully preserves the integrity of philosophical arguments, making them more discoverable than arbitrary text splits.

\end{appendices}


\end{document}